\documentclass[prd, floatfix, amsfonts, amssymb, amsmath, preprint, showkeys, nofootinbib,superscriptaddress]{revtex4-2}
\usepackage{amsmath,color}
\usepackage{multirow}
\usepackage{comment}
\usepackage{gensymb}
\usepackage{subcaption}
\usepackage{placeins}
\usepackage{caption}
\usepackage{subcaption}
\usepackage[dvipsnames, usenames]{xcolor}
\definecolor{linkcolor}{rgb}{0.0,0.3,0.5}
\usepackage[unicode, colorlinks=true, linkcolor=linkcolor, citecolor=linkcolor, filecolor=linkcolor, urlcolor=linkcolor, linktocpage, breaklinks]{hyperref}
\usepackage[all]{hypcap}
\usepackage[T1]{fontenc}
\usepackage[utf8]{inputenc}
\usepackage{mathrsfs}
\usepackage[usenames,dvipsnames]{xcolor}
\hypersetup{colorlinks=true,citecolor=romared,linkcolor=romared,urlcolor=romared}

\definecolor{romared}{RGB}{142,0,28}

\usepackage[normalem]{ulem}

\usepackage{dcolumn}
\usepackage{hyphenat}
\usepackage{graphicx}
\usepackage{ulem}
\usepackage{aas_macros}

\usepackage{orcidlink}

\begin{document}

\title{High-Density Sound Speed and Post-Merger Dynamics}

\author{John Stroud}
\affiliation{Institute for Nuclear Theory, University of Washington, Seattle, WA USA}
\email{jstroud3@uw.edu}

\author{David Radice}
\affiliation{Institute for Gravitation and The Cosmos,The Pennsylvania State University, University Park PA 16802, USA}
\affiliation{Department of Physics, The Pennsylvania State University, University Park PA 16802, USA}
\affiliation{Department of Astronomy and Astrophysics,The Pennsylvania State University, University Park PA 16802}
\email{dur555@psu.edu}
\author{Sanjay Reddy}
\affiliation{Institute for Nuclear Theory, University of Washington, Seattle, WA USA}
\email{sareddy@uw.edu}

\begin{abstract}
The density dependence of the speed of sound in cold neutron star matter remains poorly constrained and is central to determining the high-density equation of state (EOS). While binary neutron star (BNS) merger simulations increasingly incorporate detailed microphysics, the direct impact of the sound-speed in the neutron star core on observable signatures has not been systematically explored. We address this by introducing a simplified parametrization that suppresses microphysics while allowing controlled variation of the sound speed at supranuclear densities through its derivative with respect to the baryon chemical potential. Using the \text{WhiskyTHC} code, we perform a suite of fully relativistic BNS merger simulations and identify correlations between the sound-speed slope parameter and key merger outcomes, including remnant properties and post-merger GW frequencies. These results demonstrate that multi-messenger observables are sensitive to the behavior of matter at the highest densities reached in neutron star cores. We further analyze gravitational-wave signals from the CoRe database of binary neutron star merger simulations employing more realistic equations of state. Our analysis reveals approximately EOS-independent relations between the derivative of the sound speed and the peak post-merger gravitational-wave frequency. Although these relations cannot be considered truly quasi-universal, they nonetheless indicate that post-merger gravitational waves retain measurable information about the EOS at the highest densities. At the same time, the remaining EOS dependence highlights the difficulty of isolating the underlying high-density physics and motivates the development of targeted parameterized frameworks for interpreting future multi-messenger observations.
\end{abstract}
\maketitle
\section{Introduction}
\label{sec:intro}  

Neutron stars provide a direct link between nuclear physics and astrophysical observations, probing the behavior of matter at densities exceeding $n_0\approx 2.6 \times 10^{14}$ g/cm$^3$, the  nuclear saturation density, and thus constraining the high-density equation of state (EOS)~\cite{Lattimer:2012nd,Oertel:2016bki}. 

Radio, x-ray and gravitational wave (GW) observations of neutron stars during the past decade have provided valuable constraints on their masses and radii. Taken together, the existence of neutron stars with masses near or above $2~M_\odot$ \cite{Demorest:2010bx,Antoniadis:2013pzd,Fonseca:2021wxt}, the relatively small radii inferred by \textit{NICER} \cite{Riley:2019yda,Miller:2019cac,Riley:2021pdl,Miller:2021qha}, and the low tidal deformability extracted from GW170817 ~\cite{Abbott:2018exr} now point toward an EOS that is comparatively soft at $\sim(1\!-\!2)n_0$ but stiffens rapidly at higher densities in order to support the most massive stars. In phenomenological terms, this implies a marked increase of the sound speed in the inner core, often exceeding the conformal value $c_s^2=1/3$ over some density interval, and has been interpreted as evidence for strong non-perturbative effects that can produce stiffening at densities encountered in the core~\cite{Bedaque:2014sqa,Tews:2018iwm,Annala:2019puf,Ecker:2022xxx}.

In binary systems, these objects emit gravitational radiation as they inspiral, a prediction of general relativity confirmed through binary pulsar timing~\cite{Hulse:1974eb,Taylor:1982zz}, culminating in mergers that generate detectable GW signals~\cite{Abbott:2017vwq}. During the late stages of the inspiral, finite-size effects enter the waveform phase evolution through the tidal deformability parameter $\Lambda$, which depends sensitively on the stellar compactness and thus on the pressure at densities $\sim (1\text{--}2)n_0$, providing direct constraints on the intermediate-density EOS~\cite{Flanagan:2007ix,Hinderer:2007mb}. These events are accompanied by electromagnetic counterparts powered by relativistic ejecta and $r$-process nucleosynthesis, establishing binary neutron star mergers as primary sites of heavy element production~\cite{Metzger:2016pju,Kasen:2017sxr}. 

The post-merger phase probes even more extreme conditions, with densities and temperatures significantly exceeding those of cold, isolated neutron stars, and encodes additional EOS information in the GW spectrum. In particular, the dominant post-merger oscillation frequency $f_{\rm pk}$ (also referred to as $f_2$ in the literature) and related spectral features correlate with the radius and compactness of the remnant and therefore with the pressure at several times nuclear saturation density~\cite{Bauswein:2011tp,Hotokezaka:2013iia}. Similarly, the threshold mass for prompt collapse to a black hole, $M_{\rm th}$, provides a complementary constraint on the maximum mass and stiffness of the EOS~\cite{Bauswein:2013jpa,Koeppel:2019pys}. Electromagnetic observables arising from the merger ejecta provide an additional probe: the ejecta mass $M_{\rm ej}$, velocity $v_{\rm ej}$, and composition depend on the merger dynamics, including shock heating, tidal torques, and the lifetime of the hyper-massive remnant, all of which are governed by the EOS. In particular, the high-density behavior of the speed of sound influences the remnant structure and stability, thereby affecting both the efficiency of mass ejection and the kinetic energy of the outflows, which are directly imprinted in kilonova light curves and spectra. As next-generation detectors extend sensitivity into the kilohertz regime, these multi-messenger observables will provide increasingly stringent constraints on dense matter, motivating a systematic investigation of how extreme variations in the sound speed at supranuclear densities impact merger outcomes. To this end, we introduce a parametrized description of the high-density EOS in which the behavior of the speed of sound is directly controlled, enabling a targeted exploration of its influence on merger dynamics and multi-messenger observables.
\noindent
Most existing studies of binary neutron star mergers emphasize increasingly sophisticated microphysical modeling, motivated by the high computational cost of fully relativistic simulations and the goal of achieving maximal realism. In contrast, the approach adopted here is to isolate the impact of specific features of the high-density equation of state (EOS) through a simplified, controlled parametrization. In section II, we introduce a model in which the behavior of the speed of sound is governed by a single parameter, $c_s^2(\mu)'$, allowing for systematic variation of the EOS stiffness at supranuclear densities. Using this framework, we perform a suite of equal-mass neutron star merger simulations within numerical relativity, employing a $\Gamma$-law prescription for the thermal component of the EOS. 
\noindent
Within this controlled setting, in section III, we identify clear correlations between the sound-speed parameter and both GW observables and properties of the dynamically ejected material, including the ejecta mass and velocity. This study represents a first step toward a reduced-parameter description of merger outcomes, enabling a more transparent mapping between EOS properties and multi-messenger signals. We also discuss how including knowledge of the sound speed can improve fits for existing quasi-universal relationships utilizing the CoRe database . In section IV,  we conclude and comment on how parametrized models may be used to train surrogate models, including neural-network-based emulators, that aid in the extraction of EOS constraints from current and future observations by enabling efficient exploration of high-dimensional parameter spaces associated with dense matter thermodynamics.

\section{Methods and Setups}
\subsection{High Density Equation of State}
\noindent
Having outlined the broader motivation in the Introduction, we now focus on the theoretical landscape governing the equation of state (EOS) and sound speed in neutron-rich matter.A first-principles determination of the EOS across the full density range relevant to neutron stars remains an outstanding challenge. In the cores of massive neutron stars, densities may reach $6$--$8\,n_0$, where many-body correlations, higher-order interactions, and potentially new degrees of freedom become essential. At present, a fully controlled treatment of such effects is not achievable. Moreover, neutron star matter---characterized by extreme isospin asymmetry with only $\sim 5$--$10\%$ protons---cannot be directly probed in terrestrial experiments~\cite{Koehn:2024set,Lattimer:2000nx,Lattimer:2004pg}. Consequently, astrophysical observations provide the primary empirical constraints on the high-density EOS.

A central quantity derived from the EOS is the adiabatic sound speed,
\begin{equation}
    c_s^2 = \left(\frac{\partial P}{\partial \varepsilon}\right)_s \, ,
\end{equation}
which governs the propagation of linear perturbations in relativistic hydrodynamics~\cite{Rezzolla:2013rehy}. Its behavior directly impacts neutron star structure: larger values correspond to a stiffer EOS and support higher maximum masses and larger radii, while smaller values yield softer EOSs and more compact stars.

\medskip

\noindent
\textbf{Controlled regime at low densities.}
A key theoretical anchor is provided by the regime $n \lesssim 1.5\,n_{0}$, where the EOS of neutron-rich matter is well constrained by \emph{ab initio} nuclear theory. In this density range, calculations based on chiral effective field theory (EFT), incorporating consistent two- and three-nucleon interactions, exhibit a high degree of robustness. 
\noindent
Across a variety of many-body methods---including many-body perturbation theory, self-consistent Green’s functions, coupled cluster approaches, and quantum Monte Carlo simulations---predictions for the EOS and derived quantities such as the sound speed show remarkable consistency within controlled uncertainties. Early work by Hebeler and Schwenk demonstrated that chiral EFT interactions lead to convergent and systematically improvable results in neutron matter~\cite{Hebeler:2010xb}. This picture has since been reinforced by studies from Tews \emph{et al.}, Drischler \emph{et al.}, and Gandolfi, Carlson, and Reddy, which find agreement across different interaction models and computational frameworks~\cite{Tews:2012fj,Drischler:2016djf,Gandolfi:2011xu}. 
\noindent
This convergence provides evidence that the EOS and sound speed below $\sim 1.5\,n_{0}$ is known with well understood errors, establishing a solid baseline for extrapolations to higher densities where theoretical uncertainties grow rapidly.

\medskip

\medskip

\noindent
\textbf{Asymptotic regime and integral constraints.}
At asymptotically high densities, QCD becomes weakly coupled due to asymptotic freedom, and perturbative calculations become applicable. In this regime, approximate conformal symmetry is restored, implying that the sound speed approaches
\begin{equation}
    c_s^2 \to \frac{1}{3} \, c^2 \, .
\end{equation}
However, the densities at which perturbative QCD (pQCD) becomes quantitatively reliable lie far beyond those realized in neutron star interiors, typically $n_{\rm pQCD} \gtrsim 40\,n_0$~\cite{Gross:2005kv,Somasundaram:2022ztm}.
\noindent
Nevertheless, it has become increasingly clear that pQCD can indirectly constrain the EOS at much lower densities relevant to neutron stars. These constraints emerge from enforcing thermodynamic consistency when connecting the well-controlled low-density regime to the asymptotic high-density limit. In particular, integral relations between the pressure and energy density impose nonlocal constraints on the EOS, correlating the behavior of the sound speed across widely separated density scales~\cite{Annala:2019puf,Annala:2022sdu,Komoltsev:2022tsh}.
\noindent
A key insight of recent work, including that of Komoltsev and Kurkela, is that these integral constraints can significantly restrict the allowed parameter space of EOS models, even at densities well below the nominal range of validity of pQCD. Physically, this reflects the fact that excessively large values of the sound speed over an extended density interval would lead to an over-accumulation of pressure relative to the asymptotic limit, thereby violating the requirement that the EOS smoothly approach the conformal regime at high densities.
\noindent
Importantly, the impact of these constraints is highly non-uniform. They primarily exclude extreme EOS realizations characterized by very large sound speeds sustained over broad density ranges. For more moderate sound speed profiles, the resulting restrictions are comparatively mild, leaving substantial freedom in the intermediate-density regime relevant for neutron stars. We shall return to the impact of these constraints when we discuss models characterized by $c_s^2/c^2 \simeq 1$ over an extended interval and defer the discussion of the quantitative implications to a later section.

\medskip
\noindent
\textbf{Intermediate densities and sound speed behavior.}
The behavior of the sound speed in the intermediate density regime which remains most uncertain is the subject of this study. While much of the existing literature has focused on cold, static neutron star configurations, dynamical scenarios such as binary neutron star mergers provide complementary sensitivity to the microphysics of dense matter~\cite{Hotokezaka:2011dh,Shibata:2005ss}.  In particular, although the scenario in which $c_s^2$ approaches the perturbative limit monotonically is disfavored ~\cite{Bedaque:2014sqa,Tews:2018iwm} the behavior inside neutron stars, and the peak value at high density are not known. This density interval will be the focus of our study and we shall implement a range of models discussed in the next subsection. We investigate how different parameterizations of the sound speed, $c_s^2(\mu)$, impact the dynamics and observable outcomes of binary neutron star mergers, with particular emphasis on the role of intermediate-density behavior beyond the well-constrained \emph{ab initio} regime.  
\noindent
\subsection{EOS Models}
\noindent
\medskip
Following the discussion above, we now introduce the cold equations of state employed throughout this work. Up to nuclear saturation density, $n_0$, where the EOS is relatively well constrained, we adopt a realistic tabulated nuclear EOS based on the Brueckner--Hartree--Fock approximation informed by chiral perturbation theory, obtained from the CompOSE online EOS database \cite{Bombaci:2018ksa,Douchin:2001sv,Typel:2013rza,Oertel:2016bki,CompOSECoreTeam:2022ddl}. At these densities, a variety of modern microscopic approaches predict qualitatively similar behavior when the underlying two-body interactions are constrained by nucleon--nucleon scattering phase shifts and the three-body sector is calibrated to the structure and binding energies of light nuclei. In particular, many-body perturbation theory calculations based on chiral effective field theory interactions \cite{Hebeler:2009iv,Tews:2012fj,Drischler:2019wtt,Huth:2021xxh,Drischler:2020vpf} and Quantum Monte Carlo studies of neutron matter \cite{Gandolfi:2011xu,Gezerlis:2013ipa,Lynn:2015jua,Lonardoni:2019ypa,Kruger:2013kua} generally produce pressure and energy-density bands consistent with the low-density EOS employed here, lending additional support to the robustness of the EOS below and moderately above saturation density . These calculations, despite employing different many-body techniques and regulator choices, broadly agree over the density range most relevant for the outer core of neutron stars, suggesting that the dominant uncertainties arise primarily at higher densities where extrapolations beyond the regime directly constrained by nuclear data become unavoidable.

At higher densities, where theoretical uncertainties become significant, we parametrize the EOS through the behavior of the sound speed. Our goal is to investigate how the density dependence of the sound speed, and in particular the rate at which it stiffens with increasing density, influences neutron-star merger observables. To this end, we consider a simple one-parameter model in which the squared sound speed varies linearly with baryon chemical potential,
\begin{equation}
c_s^2(\mu)=c_{sm}^2 + c_{\rm ref}^{2}\frac{(\mu-\mu_m)}{\mu_m}\,,
\end{equation}
where $c_{sm}^2 \equiv c_s^2(\mu_m)$ is the sound speed at the matching point $\mu_m$, and $c_{\rm ref}^{2}$ is a measure of the slope parameter.

Given this ansatz, the remaining thermodynamic quantities follow directly from standard zero-temperature thermodynamic relations. At $T=0$ for matter in beta equilibrium, the adiabatic sound speed
\begin{equation}
c_s^2 = \left(\frac{\partial P}{\partial \varepsilon}\right)_s,
\end{equation}
together with $dP = n d\mu$  and $d\varepsilon = \mu dn$, where $n$ is the baryon number density, implies
\begin{equation}
\frac{dn}{d\mu} = \frac{n}{\mu c_s^2\left(\mu\right)}.
\end{equation}
Integrating from the matching point $\left(\mu_m,n_m,P_m\right)$ yields
\begin{equation}
n\left(\mu\right)=n_m
\exp\left[
\int_{\mu_m}^{\mu}
\frac{d\mu'}{\mu' c_s^2(\mu')}
\right],
\end{equation}
and the pressure follows from
\begin{equation}
P\left(\mu\right)=P_m+\int_{\mu_m}^{\mu} d\mu'~n\left(\mu'\right).
\end{equation}
The energy density is then obtained from the zero-temperature Gibbs relation,
\begin{equation}
\varepsilon = \mu n - P.
\end{equation}

\begin{figure}[h]
    \centering
    \begin{subfigure}[b]{0.48\linewidth}        
    \includegraphics[width=\linewidth]{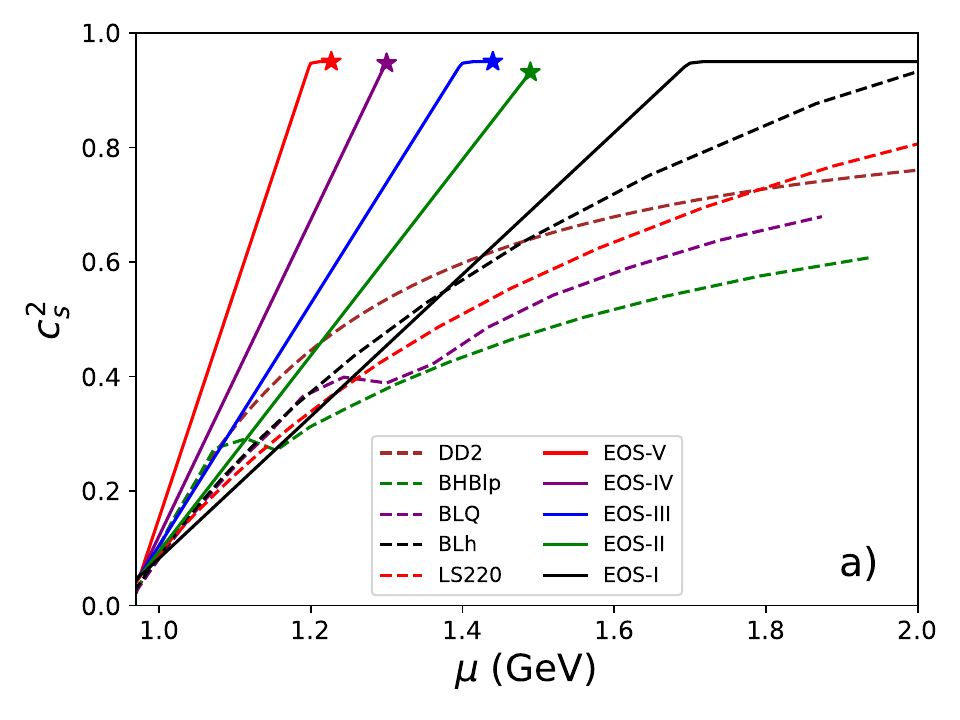}
    \caption{}
    \label{eos_model:cs2_vs_n}
    \end{subfigure}
    \hfill
    \begin{subfigure}[b]{0.48\linewidth}
        \centering
        \includegraphics[width=\linewidth]{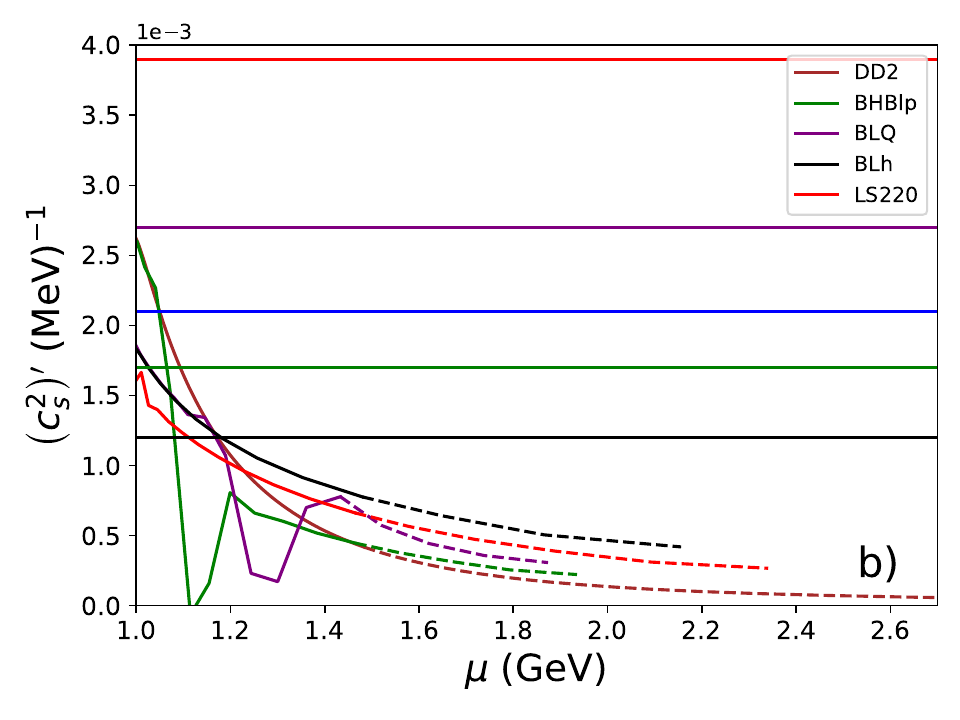}
        \caption{}
    \label{fig:dcs2_dmub_fig}
    \end{subfigure}
    \caption{(a) The speed of sound ansatz employed for our models along with a comparison with several EOSs which incorporate realistic microphysics. (b) Derivative of $c_s^2(\mu)$ as a function of $n/n_0$ for the realistic EOSs. Dotted lines show the chosen values of the parameter $(c_s^2)'$ for models $I-V$.  The solid curves are plotted up to the maximum chemical potential in our simulations.}
    \label{fig:eos_model}
\end{figure}

\begin{figure}[h!]
    \centering
    \includegraphics[width=0.6\linewidth]{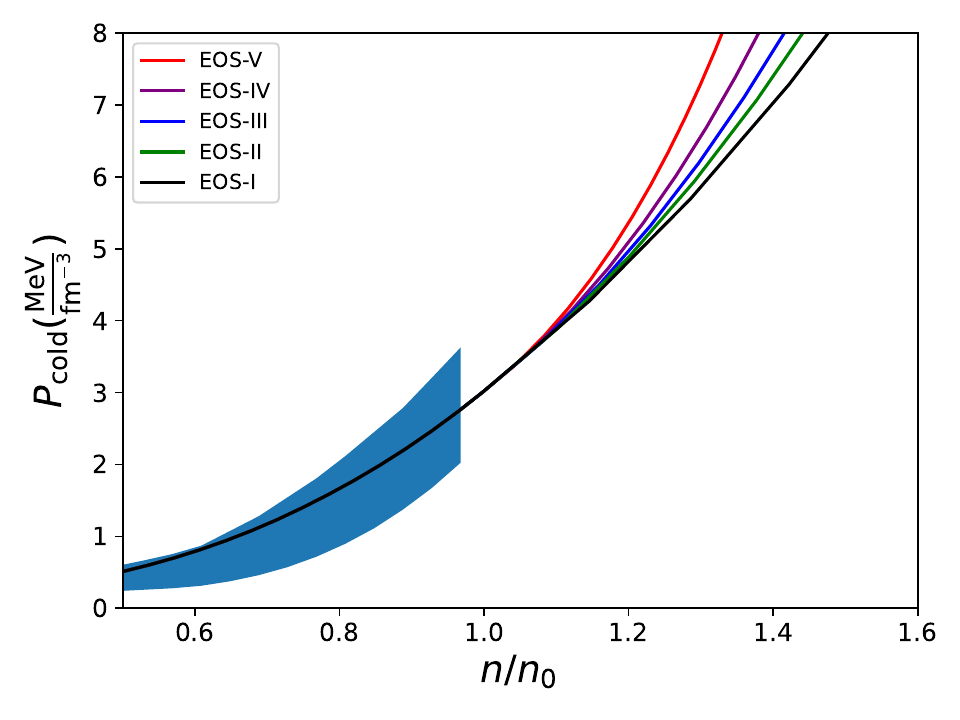}
    \caption{Pressure vs number density in terms of saturation density. For values below and just above saturation all of our EOSs use the same low density tabulated nuclear EOS. The blue region represents the allowed region based on the maximal region bounded by the CoRE equations of state}
    \label{fig:eos_pressure_density}
\end{figure}

For the linear sound-speed model considered here, the baryon density can be written analytically as
\begin{equation}
n\left(\mu\right)=n_m
\left[
\frac{\mu\, c_{sm}^2}
{\mu_m\, c_s^2\left(\mu\right)}
\right]^{
\frac{1}{c_{sm}^2-c_{\rm ref}^2}
}
\end{equation}
The sound speed is evolved until it reaches a maximum value of $c_s^2 = 0.95$. Consequently, all EOSs considered in this work exhibit a prolonged regime with large sound speed, substantially exceeding the conformal value $c_s^2 = 1/3$. Such rapid increases in the sound speed at densities relevant to neutron-star interiors must also be checked for consistency with recently derived integral constraints on the EOS \cite{Komoltsev:2022tsh}. These constraints provide additional nonlocal conditions on the permissible behavior of the sound speed and place important restrictions on strongly stiffening EOSs. We defer a detailed discussion of the compatibility of our parameterized EOSs with these integral constraints to a later section. The speed of sound as well as its derivative are shown for our models and more realistic EOSs are shown in Figure \ref{fig:eos_model}. The maximum chemical potentials which were probed in the remnant during our simulations are denoted with stars. For the softest equation of state every simulation results in a prompt collapse to a BH and so these simulations do not probe the high density remnant object. Figure \ref{fig:eos_pressure_density} shows the pressure as a function of baryon density near saturation density. The blue band shows the maximal range of pressures spanned by several equations of state with realistic microphysics. Below the matching point at saturation our EOS agrees within the range of low density pressures.  


\begin{figure} [h!]
    \centering
    \begin{subfigure}[b]{0.48\linewidth}
        \includegraphics[width=\linewidth]{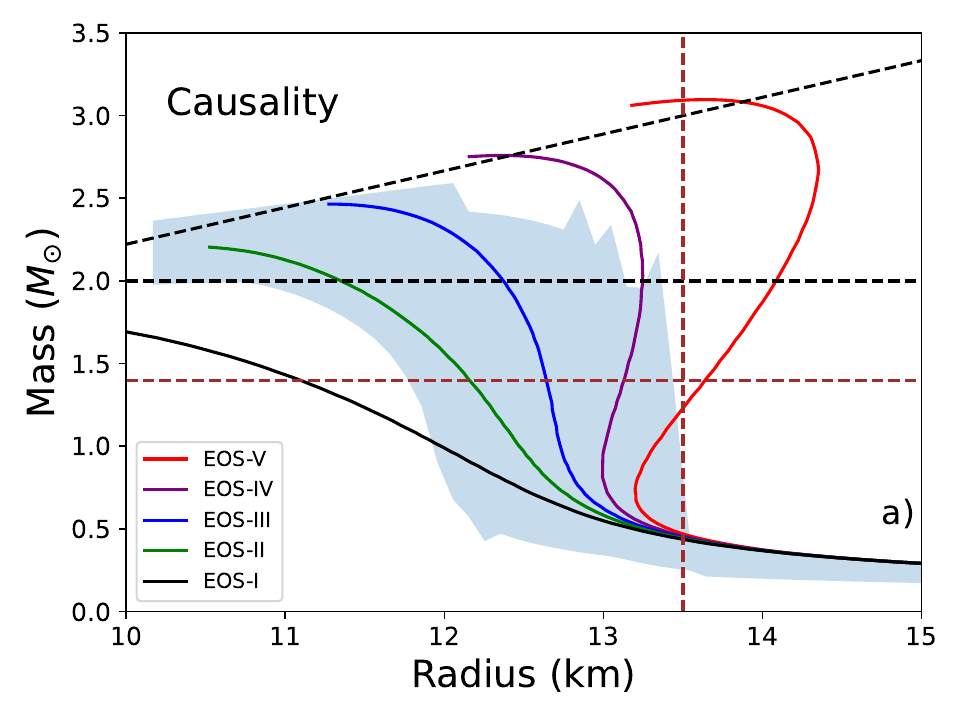}
        \caption{}
        \label{mvr_fig:a}
    \end{subfigure}
    \hfill
    \begin{subfigure}[b]{0.48\linewidth}
        \includegraphics[width=\linewidth]{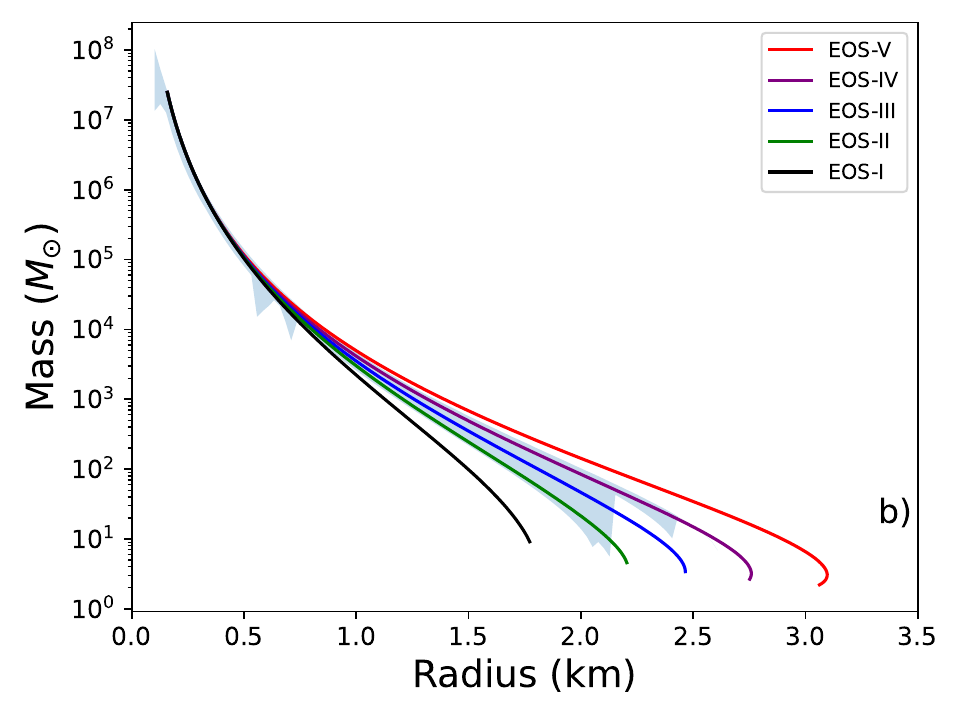}
        \caption{}
        \label{mvr_fig:b}
    \end{subfigure}
    \caption{(a) Mass vs Radius curve generated for each model along with a sample of realistic equations of state from the CoRe database as a comparison. The vertical dotted line denotes the GW radius constraint and in order for an EOS to satisfy this constraint the 1.4 $M_\odot$ star must fall to the left of this line (b) Dimensionless tidal deformability as a function of mass for the five models considered along with a comparison with realistic EOSs}
    \label{fig:mvr_lambda}
\end{figure}

\noindent
\textbf{Equation of State Constraints}
To restrict the range of possible behavior of the speed of sound at intermediate density we impose astrophysical constraints on neutron star structure.  As discussed before the maximum mass ($M_{\text{max}}$ or $M_{\text{TOV}})$ is currently unknown but observations of pulsars suggest that $M_{\text{TOV}} \geq 2.0 M_\odot$. In addition measurements of the tidal coupling in the BNS merger GW170817 was able to place upper bounds on the radius of a canonical 1.4 $M_\odot$ neutron star as $R_{1.4} \leq 13.5 \text{ km}$. Since the model we are using is quite simple we choose these two constraints to establish an acceptable range of values for the slope parameter. Using the TOV equations to solve for the structure of an isolated neutron star using our cold EOS models we can construct the mass-radius relation for our equations of state as shown in Figure \ref{fig:mvr_lambda}.  The horizontal line at $2.0~M_\odot$ represents the constraint that the maximum mass $M_{\text{TOV}} \geq 2.0~ M_\odot$. While the vertical dotted line represents the maximum radius of a 1.4 $M_\odot$ neutron star. We also compute $\Lambda$ the dimensionless tidal deformability defined by $\Lambda = \frac{2}{3}k_2C^{-5}$ where $k_2$ is the dimensionless love number determined by solving a system of differential equations derived from small perturbations of the TOV solution from a static external gravity field (in the case of a merger this would come from the other star in the binary) \cite{Hinderer:2009ca}.  We determine this quantity as the tidal deformability influences the late inspiral and is imprinted in the resulting GW emission\cite{Gonzalez:2022mgo}.    
As mentioned previously, perturbative QCD (pQCD) calculations of the equation of state at asymptotically high densities, when combined with thermodynamic consistency and stability, provide important constraints on the EOS at intermediate densities \cite{Komoltsev:2022tsh}. In particular, pQCD calculations yield quantitative estimates for the pressure and baryon number density at quark chemical potentials of order $\mu \sim 2.5$ GeV, where uncertainties associated with renormalization scale variation are sufficiently controlled to make the results useful for neutron-star EOS studies. Any viable EOS must connect smoothly to this high-density regime while respecting thermodynamic consistency, causality, and monotonicity constraints. This significantly limits the extent to which the EOS can remain stiff over a broad density interval. Sustained regions with large sound speed, $c_s^2 = \left(\partial P/\partial \epsilon\right)_s$, imply rapid growth of pressure with increasing energy density and therefore tend to overshoot the pQCD pressure band before the asymptotic regime is reached.

Komoltsev and Kurkela demonstrated quantitatively that this matching requirement strongly constrains the permissible intermediate-density behavior of the EOS, particularly models that invoke extended regions of large $c_s^2$ to generate sufficiently high pressures to support massive neutron stars while maintaining relatively small radii. A natural question is whether this tension can be alleviated by introducing a strong first-order phase transition. In principle, such a transition can soften the EOS through a discontinuous jump in energy density at nearly constant pressure, corresponding to a mixed-phase region in which $c_s^2 \simeq 0$. This can partially compensate for earlier stiffening by delaying further pressure growth and thereby permit eventual matching to the pQCD asymptote.

\begin{figure} [h!]
    \centering
    \begin{subfigure}[b]{0.48\linewidth}
        \includegraphics[width=\linewidth]{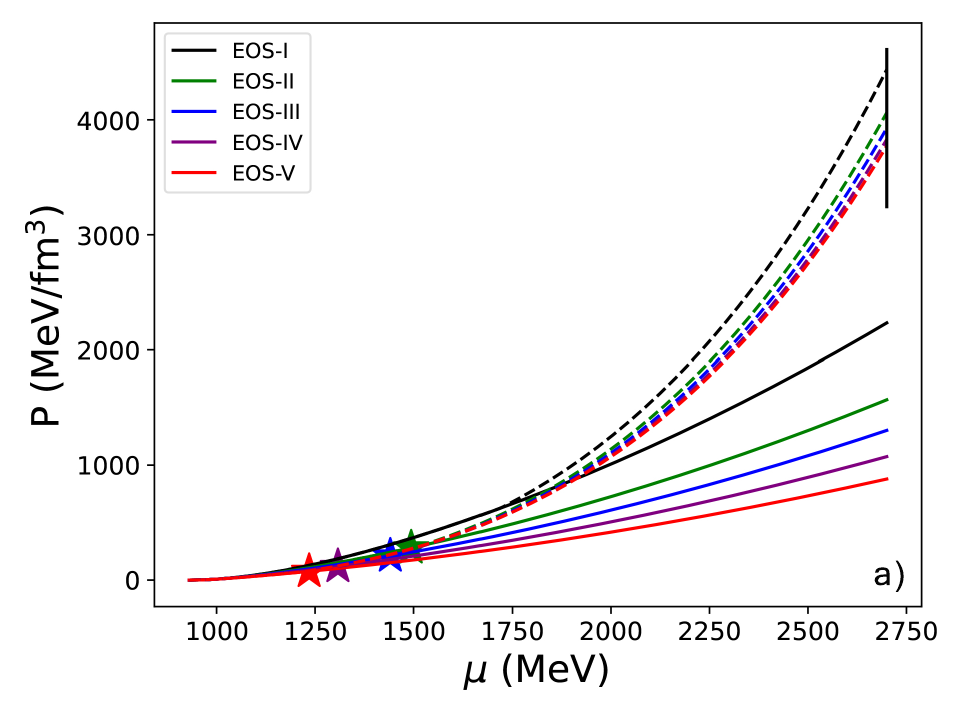}
        \caption{}
        \label{eos_ext:a}
    \end{subfigure}
    \hfill
    \begin{subfigure}[b]{0.48\linewidth}
        \includegraphics[width=\linewidth]{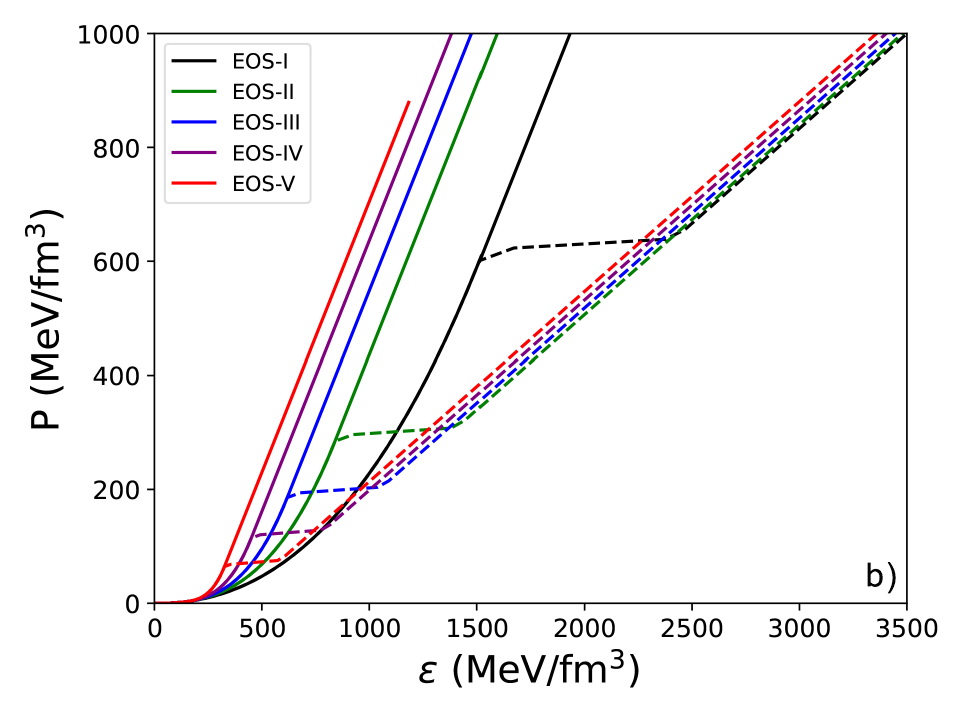}
        \caption{}
        \label{eos_ext:b}
    \end{subfigure}
    \caption{(a) Pressure vs chemical potential along with the b) equation of state for our original models denoted via solid lines as well as the models with a phase transition added at high densities as discussed in the text.}
    \label{fig:eos_extended_fig}
\end{figure}

Figure \ref{fig:eos_extended_fig}(\subref{eos_ext:a}) shows the pressure as a function of the baryon chemical potential. The perturbative QCD (pQCD) prediction at $\mu=2.7~\mathrm{GeV}$, including its estimated uncertainty, is indicated by the vertical black band using the parametrization of Ref.~\cite{Fraga:2013qra}. 

The equations of state used in our merger simulations are shown as solid curves. These models develop large sound speeds at intermediate densities and remain stiff at higher densities. Consequently, the baryon chemical potential increases rapidly with density, preventing the pressure from reaching the pQCD value at the matching point. At first sight this appears to indicate an incompatibility with the high-density pQCD constraint. However, this apparent tension should not be interpreted as a limitation of the merger simulations themselves. As shown by the stars in Figure \ref{fig:eos_extended_fig}(\subref{eos_ext:a}), our stiffest merger remnants probe maximum chemical potentials of only $\mu \lesssim 1.5~\mathrm{GeV}$, well below the $\mu=2.7~\mathrm{GeV}$ scale where the pQCD constraint is imposed. 
To demonstrate that compatibility with pQCD can nevertheless be achieved, we construct an extended family of equations of state in which the high-density behavior is modified beyond the densities relevant for mergers. Rather than remaining maximally stiff, these models undergo a first-order phase transition, producing a density interval with $c_s^2=0$. Above the transition, the sound speed approaches its asymptotic conformal value, $c_s^2=1/3$. The resulting extended equations of state are shown by the dotted curves in Figure \ref{fig:eos_extended_fig}(\subref{eos_ext:b}), while the corresponding pressure--chemical-potential relations are shown in Figure \ref{fig:eos_extended_fig}(\subref{eos_ext:a}). Unlike the original models, these extensions successfully reproduce the pQCD pressure at $\mu=2.7~\mathrm{GeV}$ while leaving the equation of state essentially unchanged throughout the density range probed by neutron star mergers.
\noindent
These results illustrate an important point. Large sound speeds inferred from binary neutron star mergers do not imply a violation of the pQCD constraint. Instead, they indicate that the sound speed must eventually decrease at densities beyond those reached in mergers. A first-order phase transition provides a simple illustrative example of such behavior, although other non-monotonic evolutions of $c_s^2$ are equally possible. 

\medskip
\noindent
\textbf{Finite Temperature}
\noindent
During a BNS merger the temperature is expected to reach $T \sim 50-100 \text{ MeV}$ where the approximation of cold matter is no longer valid. In our simulations the thermal contribution is added as an extra term to the pressure and energy density 
\begin{equation}
    P = P_{\text{cold}} + P_{\text{th}}
\end{equation}
\begin{equation}
    \varepsilon = \varepsilon_{\text{cold}} + \varepsilon_{\text{th}}
\end{equation}
\noindent
A simple yet widely used method to calculate the thermal pressure is to approximate it using an ideal gas ansatz given by 
\begin{equation}
    P_\text{th} = (\Gamma_{\text{th}} - 1)\varepsilon_{\text{th}}
\end{equation}
\noindent
Where $\Gamma_{\text{th}}$ is the ideal gas index which for an ideal gas will approach $\Gamma_{\text{th}} \to \frac{4}{3}$. As gamma is increased the equation of state becomes stiffer leading to higher thermal pressures.  Solving for $\Gamma_{\text{th}} = \frac{P_{\text{th}}}{\varepsilon_{\text{th}}} + 1$. In a realistic situation where the thermal contribution depends on local conditions one can calculate an effective ideal gas index for each particular density, temperature, and composition. Since low density neutron star matter is well described by an ideal gas of relativistic electrons the effective index approaches $\frac{4}{3}$ while for intermediate densities and temperature it varies between $1.5 \leq \Gamma \leq 2.0$, this it is only a crude approximation to assume a constant $\Gamma$ \cite{Bauswein:2010dn}.  While there are more realistic approaches which better model the density variations in $\Gamma_\text{th}$, due to the simplicity of the high density cold EOS we choose a thermal EOS with $\Gamma_\text{th} = 1.7$.

\noindent
Figure \ref{fig:dcs2_dmub_fig}  shows the derivative of $c_s^2(\mu_b)$ for realistic equations of state while the dashed lines correspond to the constant values employed in our simple model.
The speed of sound may also possess non-trivial features corresponding to possible transitions from hadronic to quark matter in NS cores, constraints from astrophysical observations and pQCD are not enough to rule out features from a crossover along with first and higher order phase transitions \cite{Mroczek:2023zxo}. Extensions of the arbitrary sound speed model may be used to constrain further features of the sound speed as a function of density which will be the subject of a future study.

\subsection{BNS Merger Simulations}
\noindent
In order to simulate BNS mergers in full general relativity one must simultaneously solve the full Einstein field equations along with the equations of general relativistic hydrodynamics (GRHD)
\begin{equation}
    G^{\mu\nu} = 8\pi T^{\mu\nu}
\end{equation}
\begin{equation}
    \nabla_\mu(T^{\mu\nu}) = 0
\end{equation}
Where $G_{\mu\nu}$ and $T_{\mu\nu}$ represent the Einstein tensor and stress-energy tensor of matter respectively.  We employ the WhiskyTHC code which is based on the Einstein Toolkit \cite{EinsteinToolkit:2025_05,Loffler:2011ay,Radice:2013xpa}. In the numerical scheme used by WhiskyTHC, The system of PDE's contained in the Einstein equations is recast using the 3 + 1 ADM formulation. This transforms the system into a Cauchy problem where the spacetime is split into slices which are spacelike hypersurfaces which can be thought of as individual snapshots of the universe at different times (see e.g. \cite{Baumgarte:2010}). The equations of GRHD are solved using a Kurganov-Tadmor scheme \cite{Kurganov:2000ovy}. the points in each consecutive slice can be connected by a vector which consists of a scalar $\alpha$ called the lapse, and a three vector $\beta^i$ known as the shift. There is an inherent gauge freedom in choosing the lapse and shift. In our simulations, the lapse is evolved with the "1 + Log" gauge condition, While the Gamma driver condition is used in evolving the shift via the CTGamma code \cite{Pollney:2009yz,Reisswig:2013sqa}.  The ADM equations themselves are split into equations which govern the evolution of the lapse and shift along with constraint equations which the relevant quantities must satisfy.  In their original form the ADM equations do not naturally lend themselves to numerical solutions. In particular a major issue in NR simulations is maintaining numerical stability and violations of the constraint equations.  The leading modern approach is a recast form of the Einstein equations known as Z4c which dampens constraint violating modes and has had much success in producing more accurate and stable simulations, our simulations utilize this approach \cite{Weyhausen:2011cg}. The WhiskyTHC code is used to simulate the final 4-6 orbits starting from initial data generated by the LORENE pseudo-spectral code which evolves the orbital separation from 100 km to 45 km \cite{Gourgoulhon:2000nn}.  The simulations utilize an adaptive mesh refinement (AMR) scheme with seven levels of refinement and a minimum resolution of $\Delta x = 245 \text{ m}$.  For simplicity we assume the system possesses reflection symmetry across the x-y plane. A low resolution system was chosen to reduce the total computational cost of the simulations.
\medskip

\noindent
\textbf{GW Extraction and Spectral Analysis}
The strain from the resulting GW is extracted via the Newmann-Penrose formalism in which the strain is calculated via the relation $\Psi_4^{lm} = \ddot{h}^{lm}_+ -  i\ddot{h}^{lm}_{\times}$ is double time integrated to determine the strain for the two GW polarizations $h_{+}$ and $h_{\times}$. Since the $l= 2,m =2$ mode is the dominant mode emitted during a BNS merger we focus solely on this one hereafter drop the $l,m$ superscripts. The Weyl scalar $\Psi_4$ is calculated from the Weyl tensor which is calculated from the trace free part of the Riemann tensor via the gauge invariant Moncrief formalism in which the perturbations are decomposed into even and odd parity modes and those are represented by spherical harmonics. From this formalism one can additionally calculate the GW luminosity $\frac{dE_{GW}}{dt}$ and radiated angular momentum $\frac{dJ_z}{dt}$ \cite{Baumgarte:2010}.  All the aforementioned analysis was performed using the \textbf{Kuibit} python package \cite{Bozzola:2021hus}. Kuibit performs the double time integration using a fixed-frequency integration in the frequency domain. Before the data is Fourier transformed a Tukey window is applied to force the waveform to zero at the endpoints.  The result of the integration is a timeseries for the two polarizations of strain. Using this we can determine amplitude of the spectral density was also calculated via $|\tilde{h}(f)|^2f^{1/2}$ where $|\tilde{h}(f)| =\sqrt{\frac{|\tilde{h}_{\times}(f)| + |\tilde{h_{+}}(f)|}{2}}$.  One prominent feature in the GW spectra is the $f_{\text{pk}}$ peak frequency which comes from the $l=2,m=2$ deformation of the remnant object which releases the most energy in GW. This peak depends on the high density matter in the remnant and has been studied for possible correlations with EOS related properties.
\medskip

\noindent
\textbf{Mass Ejecta Extraction}
In addition to the extraction of GWs from the merger we also calculate the outflow of ejected material. The ejecta across a spherical surface at ($R = 500 M_{\odot}$) using the outflow thorn of the Einstein Toolkit.  Which calculates the flux of unbound matter as well as the asymptotic velocity $v_\infty$ for each run.  The criterion for the matter to be gravitationally unbound is that $-hu_t < -1$ where $h$ is the specific enthalpy of the fluid and $u_t$ is the time component of the fluid four velocity.  
\noindent
\subsection{CoRE Database Analysis}
One of the primary goals of binary neutron star merger studies is the identification of
\emph{quasi-universal relations} (QURs) connecting measurable merger observables to the
properties of neutron stars. Such relations are termed ``quasi-universal'' because they are
largely insensitive to the microscopic details of the equation of state (EOS), depending
instead on a small number of macroscopic stellar properties such as the masses, radii,
compactnesses, or tidal deformabilities of the binary components. Consequently, QURs
provide an efficient means of inferring dense-matter properties from gravitational-wave
observations without requiring detailed knowledge of the underlying microphysics.

The best known examples relate merger observables, including the dominant post-merger
frequency, to the binary tidal deformability
$\tilde{\Lambda}(\Lambda_1,\Lambda_2)$
\cite{DelPozzo:2013ala,Gonzalez:2022mgo,Chatziioannou:2020pqz},
where the tidal deformability of an individual star is

\begin{equation}
\Lambda=\frac{2}{3}\frac{k_2}{C^5},
\end{equation}

with $k_2$ the quadrupolar Love number and
$C=M/R$ the stellar compactness.
Although remarkably successful, tidal deformability is fundamentally a
\emph{global} property of the neutron star.
It is determined by integrating the EOS throughout the stellar interior and is therefore
primarily sensitive to the pressure over the density range sampled by typical
$1.2$--$1.4\,M_\odot$ neutron stars, corresponding to characteristic densities of
approximately $2n_0$.
The central densities reached in massive neutron stars and merger remnants are
considerably higher, raising the question of whether post-merger observables retain
additional information about the EOS that is not completely encoded in macroscopic
stellar properties.

A recent development addressing this question is the IPAD--TOV (Intrinsic Perturbative Analysis of the Dimensionless TOV equations) framework~\cite{Cai:2025nxn, arXiv:2606.21402}.
Within this approach the tidal response is expressed in terms of intrinsic EOS variables,
leading to the \emph{intrinsic tidal response function}, $D(X,\Psi)$, where the dimensionless variables
\begin{equation}
X=\frac{P_c}{\varepsilon_c},
\qquad
\Psi=
2\frac{d\ln M_{\rm NS}}
{d\ln\varepsilon_c},
\end{equation}
encode information about the high-density EOS.

 The intrinsic tidal response function is the IPAD--TOV analogue of the tidal deformability. By expressing both the Love number and the compactness in terms of the intrinsic EOS variables $X$ and $\Psi$, $D$ provides a macroscopic quantity that encodes information about the integrated sound-speed profile throughout the neutron star. It therefore offers a natural framework for investigating the extent to which the microscopic behavior of the high-density EOS is captured by observable stellar properties.

In particular,
\begin{equation}
X=
\frac{1}{\varepsilon_c}
\int_0^{\varepsilon_c}
c_s^2\,d\varepsilon,
\end{equation}
is a simple average of the squared sound speed
throughout the star and represents the accumulated stiffness of the EOS from
the surface to the stellar core. Previous work has shown that combining the relation
between $\Lambda$ and $D$ with inspiral measurements can partially lift the degeneracy
between $P_c$ and $\varepsilon_c$, thereby extending EOS constraints to higher densities
than are accessible using tidal deformability alone~\cite{arXiv:2606.21402}.

Motivated by the approximately mass-independent correlation between the sound-speed
parameter and the dominant post-merger frequency found in our parameterized EOS models
(Sec.~\ref{gw_spectra_fig:mf2vsalpha}), we investigate whether information about the
high-density sound-speed profile is already encoded in these macroscopic quantities or
whether post-merger observables retain additional sensitivity to the detailed behavior
of the EOS in the stellar core.

To address this question we analyze simulations from the Computational Relativity
(CoRe) database~\cite{Dietrich:2018phi,Gonzalez:2022mgo}, which contains a large
collection of binary neutron star mergers spanning a wide range of mass ratios,
$q=m_1/m_2\ge1$, and realistic equations of state.
These include relativistic mean-field models such as Big Apple (BA),
BHB$\Lambda\phi$, and SFHo
\cite{Fattoyev:2020cws,Banik:2014qja,Steiner:2012rk},
as well as non-relativistic Skyrme models including LS220
\cite{Lattimer:1991nc}.

For binary systems we construct the intrinsic tidal response parameter

\begin{equation}
\tilde{D}(X_1,\Psi_1,X_2,\Psi_2)
=
\frac{16}{13}
\frac{(m_1+12m_2)m_1^4D_1}
{(m_1+m_2)^5}
+
(1\leftrightarrow2),
\end{equation}

which is directly analogous to the binary tidal deformability
$\tilde{\Lambda}$.
Unlike $\tilde{\Lambda}$, however,
$\tilde{D}$ depends explicitly on EOS variables that encode the integrated sound-speed
profile within the neutron star and therefore carries more direct information about
the high-density core.

Rather than searching for an entirely new quasi-universal relation, our objective is to
determine the extent to which the information contained in the microscopic sound-speed
profile is already captured by macroscopic stellar properties. We therefore examine
correlations between the post-merger peak frequency and $\tilde{D}$, and compare these
with the corresponding relations based on conventional tidal deformability.

To quantify these correlations while avoiding overfitting, we consider three simple
two-parameter models relating the dimensionless peak frequency $Mf_{\rm pk}$ to
$\tilde{D}$: a linear relation,

\begin{equation}
Mf_{\rm pk}=m\tilde{D}+b,
\end{equation}

an exponential relation,

\begin{equation}
Mf_{\rm pk}=be^{-m\tilde{D}},
\end{equation}

and a power-law relation,

\begin{equation}
Mf_{\rm pk}=b\tilde{D}^{\,m}.
\end{equation}

\section{Results}
\subsection{One Parameter Model}
\noindent
Utilizing the aforementioned methods we have performed a set of simulations with parameters and observables described in table \ref{tab:sim_table}. 
\medskip
\begin{table}[h!]
\centering
\begin{tabular}{|c|c|c|c|c|c|}
    \hline
    Model & $M_{NS}~\left(M_\odot\right)$ & $E_{GW}(M_\odot)$ & $J_{GW}~\left(M_\odot^2\right)$  & $f_{\text{pk}}\text{ (kHz)}$ & $t_{BH}~\text{(ms)}$\\
    \hline
    EOS-I  & 1.3   & 0.0256 & 1.2442  &-& 0.129\\
    \hline
    EOS-I  & 1.4   & 0.0341 & 0.7584  &-&0.046\\
    \hline
    EOS-II  & 1.3  & 0.0623 & 1.9006  & 3.135 &-\\
    \hline 
    EOS-II  & 1.4  & 0.0446& 1.6078  & - & 0.513\\
    \hline
    EOS-III  & 1.3 & 0.04927 & 1.6704  & 2.837 &-\\
    \hline
    EOS-III  & 1.4 & 0.0591 & 1.9442  & 2.935 &-\\
    \hline
    EOS-IV  & 1.3  & 0.0305 & 1.2514  & 2.567 &-\\
    \hline
    EOS-IV  & 1.35 & 0.0423 & 1.5703  & 2.640 & -\\
    \hline
    EOS-IV  & 1.4 &  0.0441 & 1.6408  & 2.742 & -\\
    \hline
    EOS-V  & 1.3  &  0.0274 & 1.1953 & 2.260 &-\\
    \hline
    EOS-V  & 1.4  & 0.0166 & 1.3692  & 2.318 &-\\
    \hline
\end{tabular}
\caption{Table of runs for all of the performed simulations. are the total energy and angular momentum radiated by GWs extracted at $t = 4900 \text{ M}_{\odot}$ when for all but the Model C 1.3 M saturate.}
\label{tab:sim_table}
\end{table}

\begin{figure}[h!]
    \centering
    \includegraphics[width=0.65\linewidth]{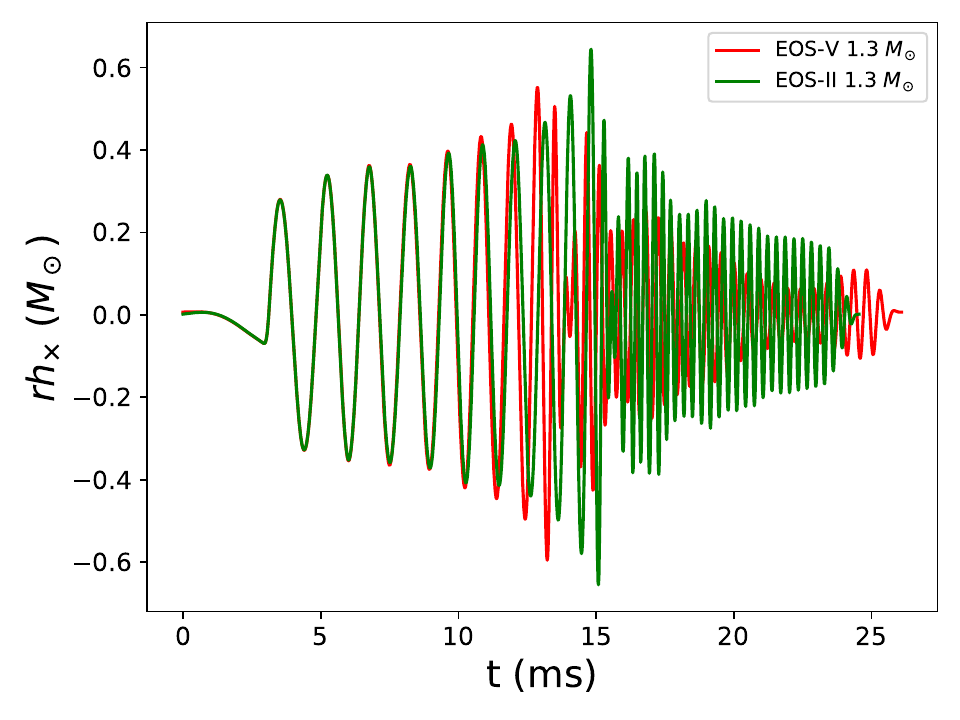}
    \caption{$h_\times$ GW polarization times the extraction radius for the softest and stiffest equations of state which don't collapse to a black hole.}
    \label{fig:rhcross_fig}
\end{figure}

\noindent
\textbf{GW Analysis}
Using the Kuibit library we extract the timeseries $rh_\times$ and $rh_+$. Figure \ref{fig:rhcross_fig} shows the cross polarizaation from a 1.3M binary from both the stiffest and softest EOS which did not result in a BH.  
\noindent
\begin{figure}[h!]
    \centering
    \includegraphics[width=0.65\linewidth]{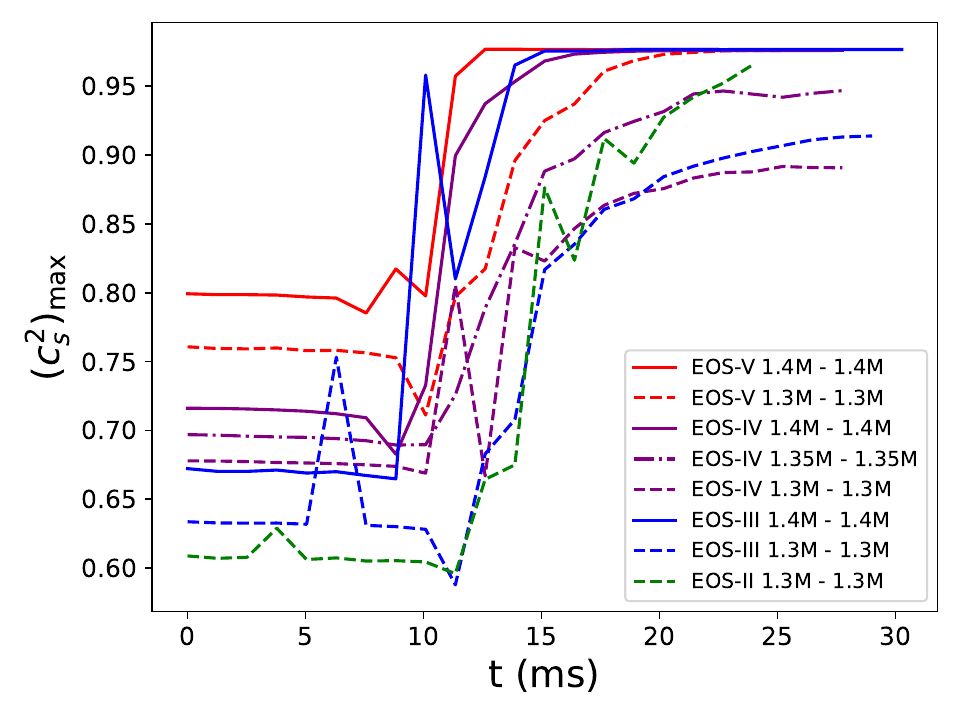}
    \caption{Maximum $c_s^2$ value as a function of simulation time for the runs which did not result in a black hole forming}
    \label{fig:bns_maxcs2}
\end{figure}

\begin{figure}[h!]
    \centering
    \begin{subfigure}[b]{0.48\linewidth}
        \includegraphics[width=\linewidth]{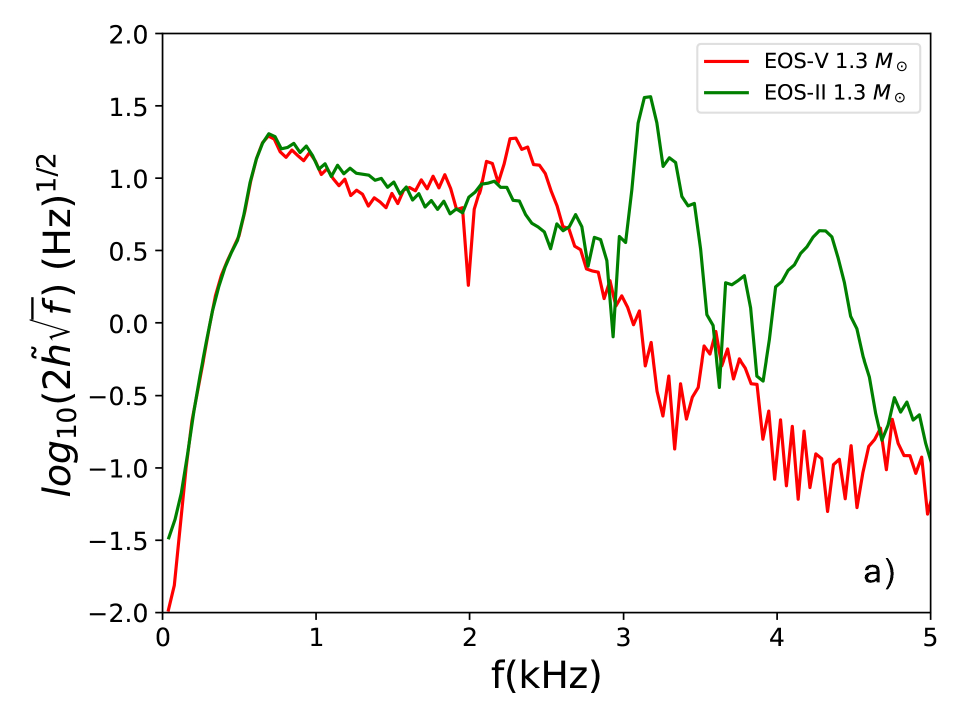}
        \caption{}
        \label{gw_spectra_fig:asd_fig}
    \end{subfigure}
    \hfill
    \begin{subfigure}[b]{0.48\linewidth}
         \centering
        \includegraphics[width=\linewidth]{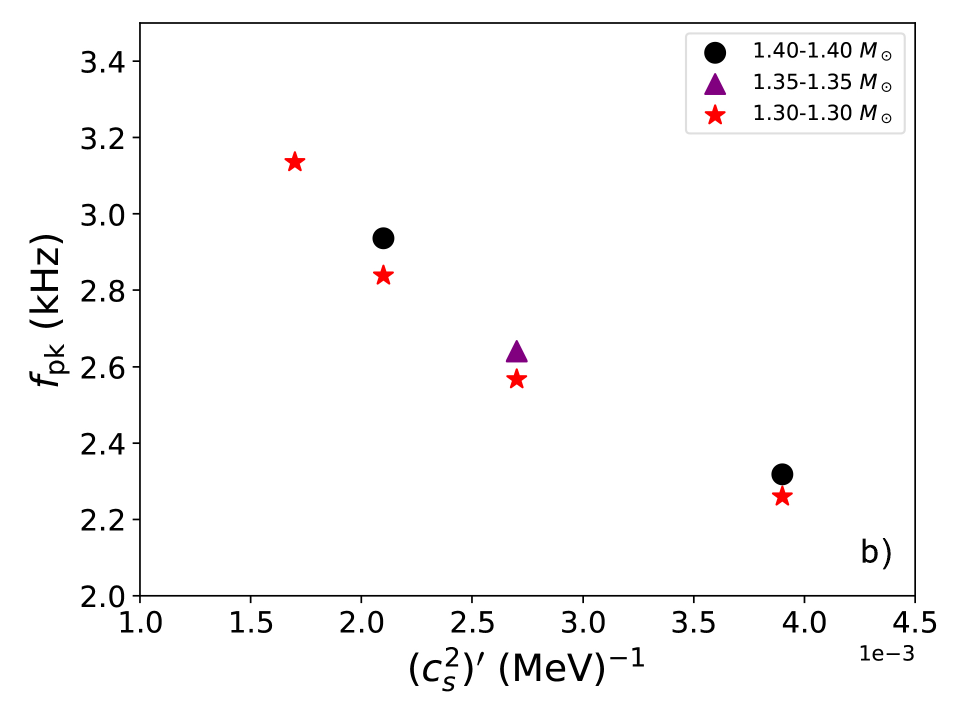}
        \caption{}
        \label{gw_spectra_fig:mf2vsalpha}
    \end{subfigure}
    \caption{a) Amplitude spectral density from GWs for the runs which did not result in a black hole forming. b)Post-merger peak frequency for each simulations as a function of the slope parameter of $c_s^2{'}$}
    \label{fig:gw_spectra_fig}
\end{figure}

Having extracted the strain we now calculate the amplitude spectral density (ASD). Figure \ref{fig:gw_spectra_fig}(\subref{gw_spectra_fig:asd_fig}) shows the ASD for the same two representative EOSs discussed above. The peak in both spectra around 2-3 kHz is the dominant post-merger peak frequency $f_\text{pk}$.

\begin{figure}[h!]
    \centering
    \begin{subfigure}[b]{0.48\linewidth}
        \centering
        \includegraphics[width=\linewidth]{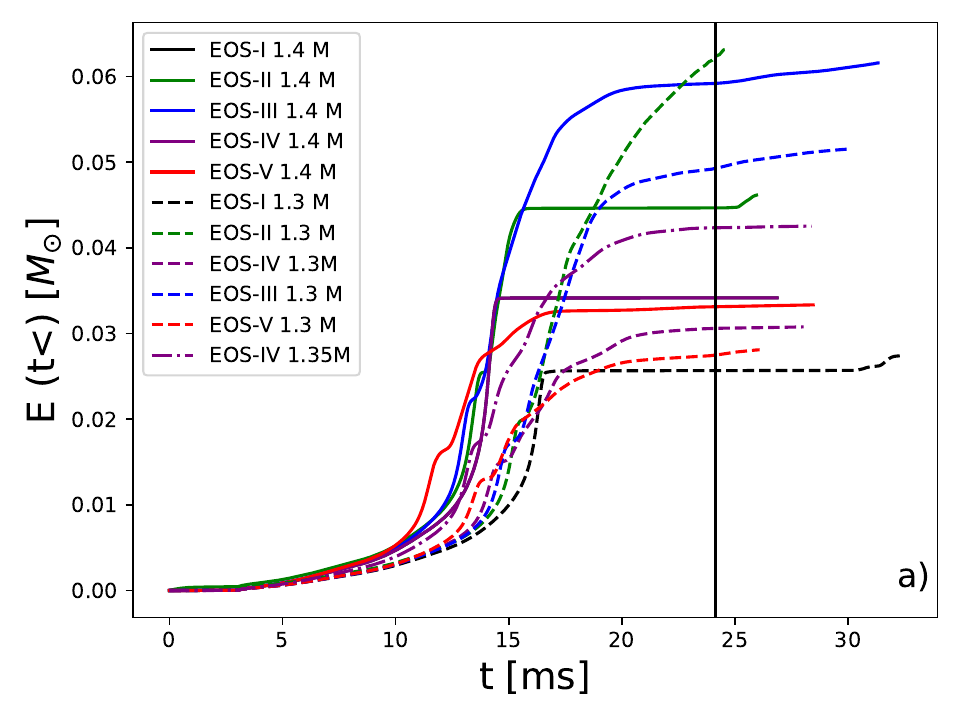}
        \label{fig:e_gw_cumulative}
    \end{subfigure}
    \hfill
    \begin{subfigure}[b]{0.48\linewidth}
        \centering
        \includegraphics[width=\linewidth]{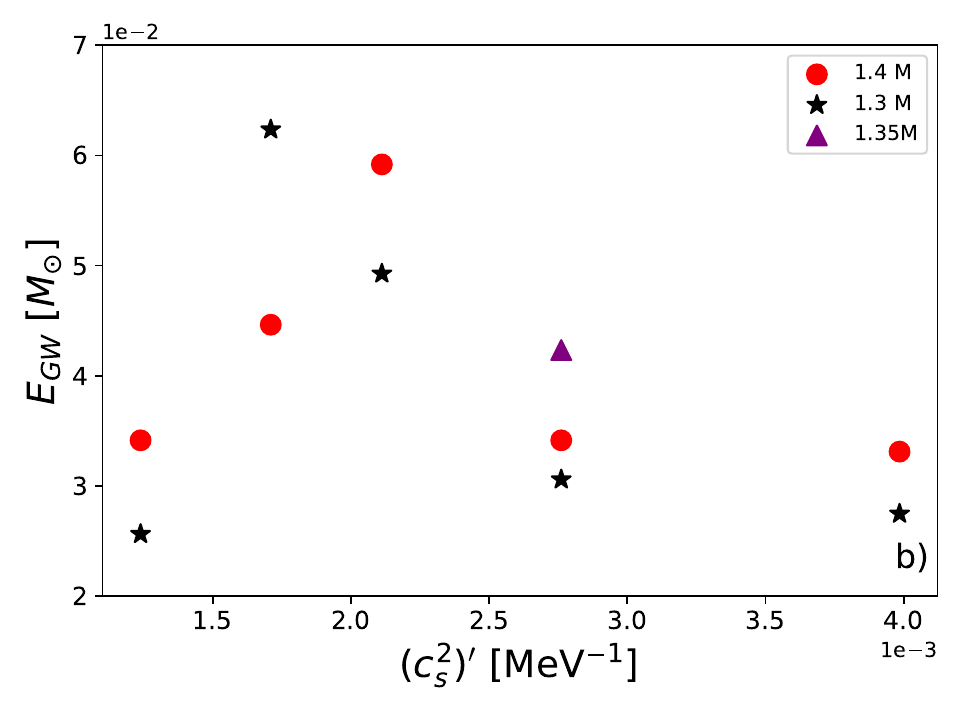}
        \label{fig:j_gw_cumulative}
    \end{subfigure}
    \caption{(a) Plot of the cumulative released energy in GWs as a function of simulation time. The vertical line denotes the point at which the total energy radiated in GW's was calculated and this data is shown in (b) as a function of the sound speed slope parameter}
    \label{fig:e_gw_j_gw_cumulative}
\end{figure}

Repeating this analysis for all of our simulations we investigated the correlations between the GW observables and our model parameter $c_s^2(\mu)'$.  
Figure \ref{fig:gw_spectra_fig}(\subref{gw_spectra_fig:mf2vsalpha}) shows $f_\text{pk}$ as a function of the slope parameter. We note that there is an approximately linear correlation between $f_{\text{pk}}$ and $c_s^2(\mu)'$ for lower values of the sound speed slope. In addition these seem to be relatively mass independent with similar values for equal mass mergers of $1.3M\odot,1.35M_\odot, \text{ and } 1.4M_\odot$ which may indicate a possible relation between $f_\text{pk}$ and the slope insensitive to the mass of the binary.      
In addition we have calculated the total energy and angular momentum radiated from GW's. As they both show similar behavior we only show the energy $E_{GW}$ in Figure \ref{fig:e_gw_j_gw_cumulative} which exhibit a distinct peaked behavior in the sound speed parameter the mechanism behind this is similar to the trend found in the mass ejected during the merger and the explanation will be postponed to that section.

\begin{figure}[h!]
    \centering
    \includegraphics[width=0.5\linewidth]{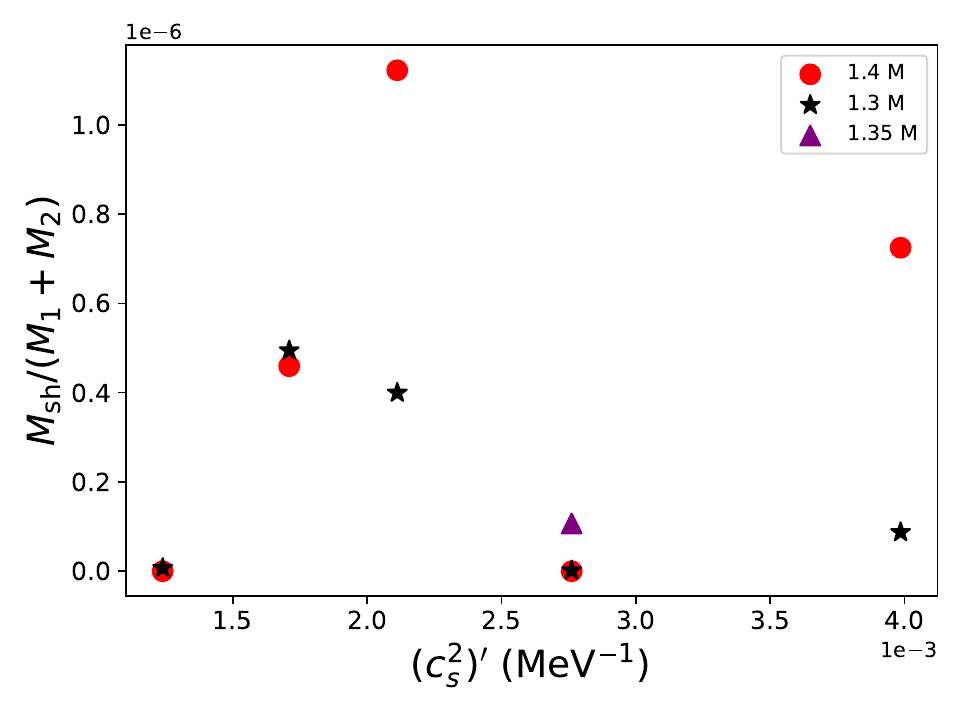}
    \caption{Mass of the shocked ejecta ($v_\infty > 0.6\text{ c}$) as a function of the slope speed parameter for simulations of 1.3M-1.3M, 1.35M - 1.35M, and 1.4M - 1.4M}
    \label{fig:m_shock_cs2prime}
\end{figure}
\begin{figure}[h!]
    \centering
    \includegraphics[width=0.8\linewidth]{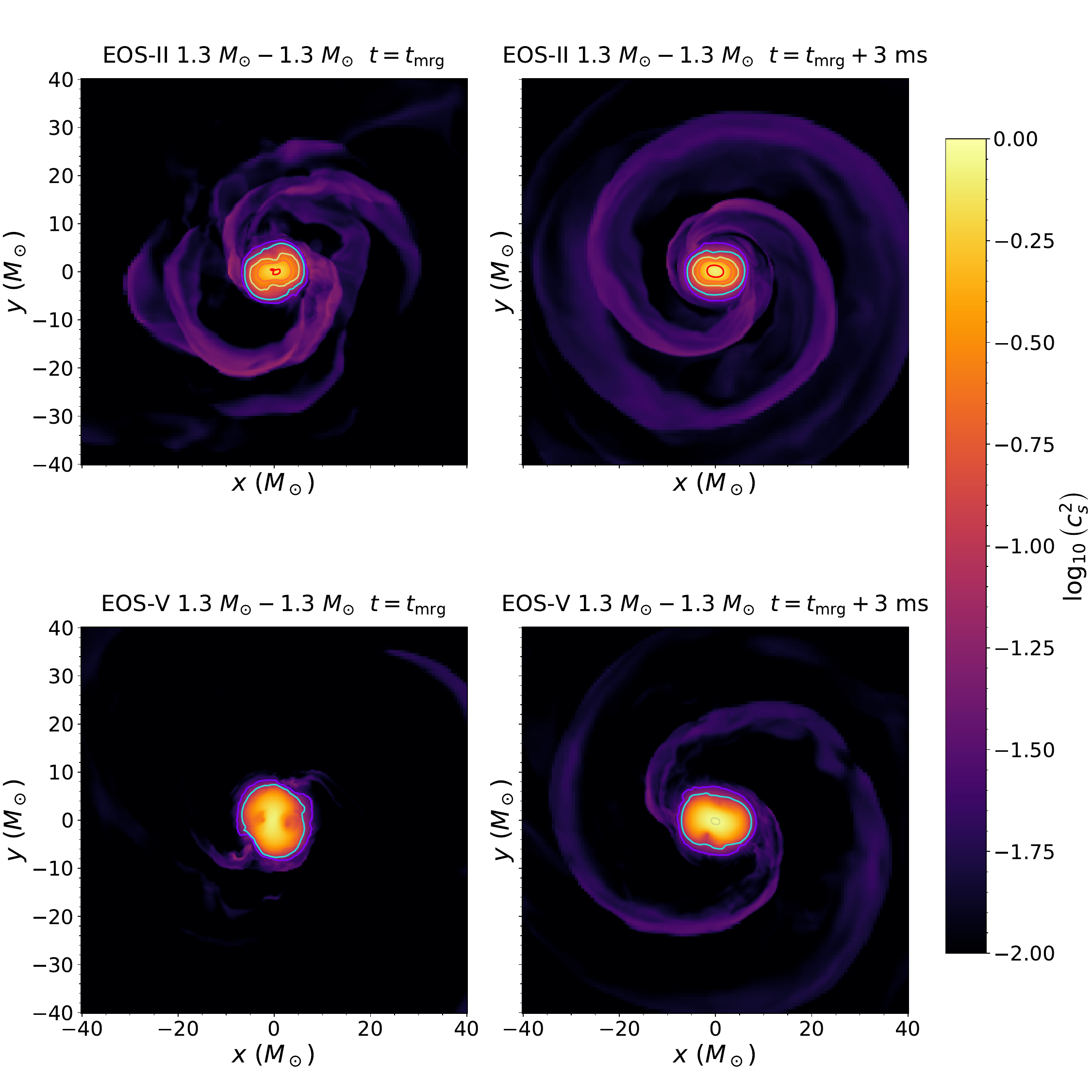}
    \caption{2D color maps of the adiabatic sound speed for the softest and stiffest EOS mergers which did not collapse in the simulation time.  Contours of constant density are plotted over the remnant at densities of $n_0/2,n_0,2n_0,3n_0,4n_0$}
    \label{fig:BNS_t_fig}
\end{figure}

\medskip
\noindent
\textbf{Ejecta Analysis}
\noindent
Next we analyze the outflow of the fast dynamically launched ejecta with asymptotic velocity $v_{\infty} > 0.6$.  The main sources for this ejecta will be from the initial interface between the two colliding stars as well as material launched by the post merger bounce \cite{Rosswog:2024vfe}.  We show the amount of mass ejected normalized by the total system mass in Figure \ref{fig:m_shock_cs2prime}.  For very small values of the slope parameter the remnant quickly collapses to a BH swallowing up much of the matter leading to very little ejected matter. At very high values of the sound speed parameter the stiffness of the remanant will likely lead to a less violent collapse launching a weaker bounce and ejecting less material. For equations of state which are soft but not so soft that they collapse the remnant would undergo a more violent collapse. The competition between these effects explain the peaked behavior we see in Figure \ref{fig:m_shock_cs2prime}. The same peaked behavior was observed for the energy and momentum released in GWs, thus there is a sort of optimal parameter which produces a soft enough equation of state so that the collision is sufficiently violent, but not so soft that the result is a black hole.  We show in Figure \ref{fig:BNS_t_fig}, a 2D color plot of the $c_s^2$ for the softest and stiffest EOSs which do not result in a BH, snapshots taken at both $t_{\text{mrg}}$ and 3ms later show, for the softer EOS a considerable amount of high density ejecta has already been thrown off tidally at the moment of merger. In addition we have overlayed contours of constant density at $n_0/2$,$n_0$,$2n_0$,$3n_0$,$4n_0$. From this we see that as expected the object in the softer case is much more compact than in the stiff case.

We have demonstrated several relationships between our model parameter and the GW spectra and mass ejecta. However, since our models are simple we should not expect these relationships to carry over to the more realistic case. In the next section we introduce our analysis of a database of more realistic merger simulations. 

\subsection{Comparison with Realistic Binary Neutron Star Mergers}

\noindent
The primary goal of the analysis is to explore the possibility of a quasi-universal relation between some measurable quantity in the merger and properties of the sound speed. A relationship is considered quasi-universal if it is more or less independent of all but some subset of the EOS parameters this relation should be more or less insensitive to the choice of EOS as well as input parameters to the binary system such as mass ratio or spin. Such relationships may provide a very clean method for determining EOS properties via merger observables \cite{Bernuzzi:2014kca}.  Employing the volume averaging procedure described in the methods section we investigated correlations between $f_{\text{pk}}$ and volume averaged values of $c_s^2$ as well as $( c_s^2(\mu)' )_{\text{avg}}$, however we were unable to find any novel universal relations between these properties of the EOS and the GW spectra. Instead, we uncovered a relation between $f_{\rm pk}$ and the recently proposed $\tilde D$ quantity, which also encodes information about the stiffness of the EOS at supernuclear densities.

\begin{figure}[!h]
    \centering
    \includegraphics[width=0.75\linewidth]{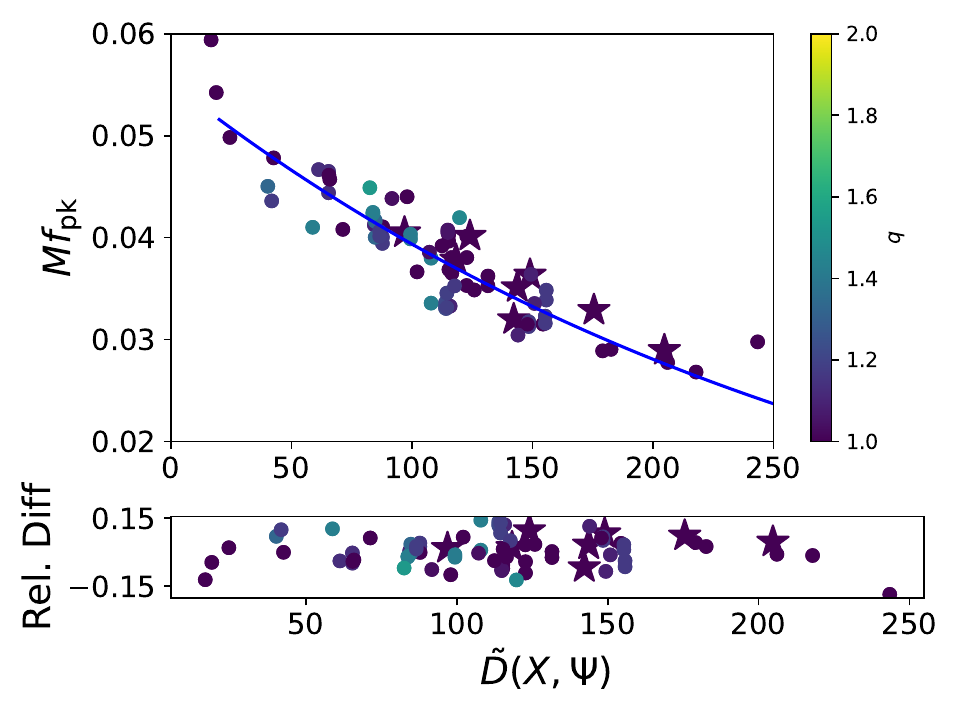}
    \caption{Quasi-Universal relation between $\tilde{D} \sim Mf_{\text{pk}}$. The color of each point represents the mass ratio q corresponding to the colorbar. Our simulations are represented as stars}
    \label{fig:mf2_qur}
\end{figure}
\noindent
\textbf{Quasi-Universal Relation} Figure \ref{fig:mf2_qur} shows the fitted quasi-universal relation as discussed in the methods section along with the relative difference between the predicted values and the observed values both from CoRe and our models.  The parameters for each fit are given in Table \ref{tab:fit_coeff} along with the standard deviations and calculated $R^2$. We find that the exponential fit performs the best of our three candidates and it is shown plotted in Figure \ref{fig:mf2_qur}. Thus a measurement of the GW spectra would provide a way to determine $\tilde{D}$ and therefore the parameter $X$ showing that the IPAD--TOV formalism can be extended to BNS merger observables. 

\begin{table}
    \centering
    \begin{tabular}{|c|c|c|c|}
    \hline
        Model & m & b & $R^2$\\
        \hline
        $m\tilde{D} + b$ & $-1.22\cdot10^{-4}\pm~7.34\cdot10^{-6}$ &$5.19\cdot10^{-2} \pm ~ 8.87\cdot10^{-4}$&0.779\\
        \hline
        $be^{-m\tilde{D}}$& $3.38\cdot10^{-3}\pm~1.85\cdot10^{-4}$ &$5.51\cdot10^{-2} \pm 1.10\cdot10^{-3}$&0.811\\
        \hline
        $b\tilde{D}^m$ & $-2.52\cdot10^{-1}\pm~1.40\cdot10^{-2}$ &$1.21\cdot10^{-1}\pm~7.70\cdot10^{-2}$&0.775\\
        \hline
    \end{tabular}
    \caption{Fit coefficients for the $Mf_{\text{pk}}$}
    \label{tab:fit_coeff}
\end{table}
\section{Conclusions}
\noindent
The density dependence of the speed of sound at supranuclear densities remains one of the
largest uncertainties in the neutron-star equation of state and plays a central role in
determining the structure of massive neutron stars and the dynamics of binary neutron-star mergers.
In this work we introduced a simple parameterized equation of state in which the
high-density behavior of the sound speed is controlled by a single parameter. This
framework suppresses much of the complexity associated with realistic microphysics,
allowing us to isolate the influence of the sound-speed profile on merger dynamics and
multi-messenger observables.

Our simulations demonstrate that the evolution of the sound speed in the stellar core has
a measurable impact on several key merger observables. In particular, the dominant
post-merger gravitational-wave frequency exhibits an approximately monotonic dependence
on the sound-speed slope (Fig.~\ref{gw_spectra_fig:mf2vsalpha}), while the gravitational-wave
energy and the mass of the shocked ejecta display a non-monotonic dependence, reflecting
the competition between prompt collapse for softer equations of state and the weaker
post-merger bounce produced by very stiff remnants (Figs.~\ref{fig:e_gw_j_gw_cumulative} and
\ref{fig:m_shock_cs2prime}). The remnant structure shown in Fig.~\ref{fig:BNS_t_fig}
provides a physical explanation for these trends by illustrating how the compactness and
high-density core evolve as the equation of state is varied.

While these results demonstrate that post-merger observables are sensitive to the
high-density sound speed, a more fundamental question is whether this microscopic
information is already encoded in the macroscopic properties of neutron stars.
Quantities such as the mass, radius, compactness, and tidal deformability are determined
by integrating the equation of state throughout the stellar interior and therefore provide
a compressed description of the underlying sound-speed profile. If these macroscopic
properties completely characterize the relevant dense-matter physics, then post-merger
observables should depend only on them, independent of the detailed behavior of the
equation of state in the core.

To investigate this question, we analyzed merger simulations from the CoRe database using
the recently developed IPAD--TOV framework~\cite{Cai:2025nxn, arXiv:2606.21402}. In this approach, the intrinsic tidal response
function $\tilde D(X,\Psi)$ is constructed from dimensionless EOS variables that encode the
integrated behavior of the sound speed within the star, with
$X=\varepsilon_c^{-1}\int c_s^2\,d\varepsilon$ representing the energy-density weighted
average sound speed. Figure~\ref{fig:mf2_qur} demonstrates that incorporating these quantities
reduces the scatter in the relation between the post-merger peak frequency
and the binary properties, indicating that a significant fraction of the microscopic
information contained in the high-density sound-speed profile is encoded in macroscopic
stellar observables.

At the same time, the remaining scatter in Fig.~\ref{fig:mf2_qur} demonstrates that this
encoding is not complete. While global quantities such as the compactness and tidal
deformability capture much of the influence of the equation of state, they do not fully
characterize the detailed behavior of matter at the highest densities reached in neutron
stars and merger remnants. Consequently, post-merger gravitational waves retain
additional information about the core equation of state beyond that accessible through
inspiral measurements alone. Rather than viewing the absence of a perfectly
quasi-universal relation as a limitation, we interpret it as evidence that post-merger
observations provide genuinely new constraints on dense matter.

The present work therefore represents a first step toward understanding how microscopic
properties of the equation of state are mapped onto observable merger signatures. The
parameterized framework introduced here provides a transparent means of isolating the role
of individual features of the sound-speed profile and can be systematically extended to
include additional physically motivated parameters, such as the density at which the sound
speed reaches its maximum, the magnitude of that maximum, and possible non-monotonic
behavior associated with phase transitions. Such reduced-order parameterizations provide
an attractive foundation for constructing surrogate models and machine-learning emulators
capable of rapidly exploring the high-dimensional EOS parameter space. As next-generation
gravitational-wave detectors begin to detect post-merger signals with increasing
sensitivity, these methods offer a promising route toward determining which aspects of
the microscopic equation of state are already encoded in macroscopic neutron-star
properties and which remain uniquely accessible through the dynamics of binary neutron
star mergers.
\section{Acknowledgments}
\noindent
DR acknowledges support from the U.S.~Department of Energy, Office of Science, Division of Nuclear Physics under Award Number(s) DE-SC0024388, and from the National Science Foundation under Grants PHY-2020275, PHY-2116686, PHY-2407681, and PHY-2512802.
The simulations in this work were performed on the Expanse cluster at SDSC through allocation PHY240157 from the Advanced Cyberstructure Coordination Ecosystem (ACCESS) program.
\bibliographystyle{apsrev4-2}
\bibliography{refs_sr,refs_js}

@article{Bozzola:2021hus,
    author = "Bozzola, Gabriele",
    title = "{kuibit: Analyzing Einstein Toolkit simulations with Python}",
    eprint = "2104.06376",
    archivePrefix = "arXiv",
    primaryClass = "gr-qc",
    doi = "10.21105/joss.03099",
    journal = "J. Open Source Softw.",
    volume = "6",
    number = "60",
    pages = "3099",
    year = "2021"
}

@article{Kurganov:2000ovy,
    author = "Kurganov, Alexander and Tadmor, Eitan",
    title = "{New High-Resolution Central Schemes for Nonlinear Conservation Laws and Convection{\textendash}Diffusion Equations}",
    doi = "10.1006/jcph.2000.6459",
    journal = "J. Comput. Phys.",
    volume = "160",
    pages = "241--282",
    year = "2000"
}

@article{Pollney:2009yz,
    author = "Pollney, Denis and Reisswig, Christian and Schnetter, Erik and Dorband, Nils and Diener, Peter",
    title = "{High accuracy binary black hole simulations with an extended wave zone}",
    eprint = "0910.3803",
    archivePrefix = "arXiv",
    primaryClass = "gr-qc",
    doi = "10.1103/PhysRevD.83.044045",
    journal = "Phys. Rev. D",
    volume = "83",
    pages = "044045",
    year = "2011"
}

@article{Reisswig:2013sqa,
    author = "Reisswig, C. and Ott, C. D. and Abdikamalov, E. and Haas, R. and Moesta, P. and Schnetter, E.",
    title = "{Formation and Coalescence of Cosmological Supermassive Black Hole Binaries in Supermassive Star Collapse}",
    eprint = "1304.7787",
    archivePrefix = "arXiv",
    primaryClass = "astro-ph.CO",
    doi = "10.1103/PhysRevLett.111.151101",
    journal = "Phys. Rev. Lett.",
    volume = "111",
    pages = "151101",
    year = "2013"
}

@article{Gourgoulhon:2000nn,
    author = "Gourgoulhon, Eric and Grandclement, Philippe and Taniguchi, Keisuke and Marck, Jean-Alain and Bonazzola, Silvano",
    title = "{Quasiequilibrium sequences of synchronized and irrotational binary neutron stars in general relativity: 1. Method and tests}",
    eprint = "gr-qc/0007028",
    archivePrefix = "arXiv",
    doi = "10.1103/PhysRevD.63.064029",
    journal = "Phys. Rev. D",
    volume = "63",
    pages = "064029",
    year = "2001"
}

@article{Koehn:2024set,
    author = "Koehn, Hauke and others",
    title = "{From existing and new nuclear and astrophysical constraints to stringent limits on the equation of state of neutron-rich dense matter}",
    eprint = "2402.04172",
    archivePrefix = "arXiv",
    primaryClass = "astro-ph.HE",
    reportNumber = "LA-UR-24-20420",
    doi = "10.1103/PhysRevX.15.021014",
    journal = "Phys. Rev. X",
    volume = "15",
    number = "2",
    pages = "021014",
    year = "2025"
}

@article{Lattimer:2000nx,
    author = "Lattimer, J. M. and Prakash, M.",
    title = "{Neutron star structure and the equation of state}",
    eprint = "astro-ph/0002232",
    archivePrefix = "arXiv",
    doi = "10.1086/319702",
    journal = "Astrophys. J.",
    volume = "550",
    pages = "426",
    year = "2001"
}

@article{Lattimer:2004pg,
    author = "Lattimer, J. M. and Prakash, M.",
    title = "{The physics of neutron stars}",
    eprint = "astro-ph/0405262",
    archivePrefix = "arXiv",
    doi = "10.1126/science.1090720",
    journal = "Science",
    volume = "304",
    number="5670",
    pages = "536--542",
    year = "2004"
}

@book{Rezzolla:2013rehy,
    author = "Rezzolla, Luciano and Zanotti, Olindo",
    title = "{Relativistic Hydrodynamics}",
    doi = "10.1093/acprof:oso/9780198528906.001.0001",
    isbn = "978-0-19-852890-6, 978-0-19-174650-5",
    publisher = "Oxford University Press",
    address = "Oxford",
    year = "2013"
}

@article{Gross:2005kv,
    author = "Gross, D. J.",
    title = "{The discovery of asymptotic freedom and the emergence of QCD}",
    doi = "10.1073/pnas.0503831102",
    journal = "Proc. Nat. Acad. Sci.",
    volume = "102",
    pages = "9099--9108",
    year = "2005"
}

@article{Somasundaram:2022ztm,
    author = "Somasundaram, Rahul and Tews, Ingo and Margueron, J{\'e}r{\^o}me",
    title = "{Perturbative QCD and the neutron star equation~of state}",
    eprint = "2204.14039",
    archivePrefix = "arXiv",
    primaryClass = "nucl-th",
    reportNumber = "LA-UR-22-22575",
    doi = "10.1103/PhysRevC.107.L052801",
    journal = "Phys. Rev. C",
    volume = "107",
    number = "5",
    pages = "L052801",
    year = "2023"
}

@article{Hotokezaka:2011dh,
    author = "Hotokezaka, Kenta and Kyutoku, Koutarou and Okawa, Hirotada and Shibata, Masaru and Kiuchi, Kenta",
    title = "{Binary Neutron Star Mergers: Dependence on the Nuclear Equation of State}",
    eprint = "1105.4370",
    archivePrefix = "arXiv",
    primaryClass = "astro-ph.HE",
    doi = "10.1103/PhysRevD.83.124008",
    journal = "Phys. Rev. D",
    volume = "83",
    pages = "124008",
    year = "2011"
}

@article{Shibata:2005ss,
    author = "Shibata, Masaru and Taniguchi, Keisuke and Uryu, Koji",
    title = "{Merger of binary neutron stars with realistic equations of state in full general relativity}",
    eprint = "gr-qc/0503119",
    archivePrefix = "arXiv",
    doi = "10.1103/PhysRevD.71.084021",
    journal = "Phys. Rev. D",
    volume = "71",
    pages = "084021",
    year = "2005"
}

@article{Bombaci:2018ksa,
    author = "Bombaci, Ignazio and Logoteta, Domenico",
    title = "{Equation of state of dense nuclear matter and neutron star structure from nuclear chiral interactions}",
    eprint = "1805.11846",
    archivePrefix = "arXiv",
    primaryClass = "astro-ph.HE",
    doi = "10.1051/0004-6361/201731604",
    journal = "Astron. Astrophys.",
    volume = "609",
    pages = "A128",
    year = "2018"
}

@article{Douchin:2001sv,
    author = "Douchin, F. and Haensel, P.",
    title = "{A unified equation of state of dense matter and neutron star structure}",
    eprint = "astro-ph/0111092",
    archivePrefix = "arXiv",
    doi = "10.1051/0004-6361:20011402",
    journal = "Astron. Astrophys.",
    volume = "380",
    pages = "151",
    year = "2001"
}

@article{Typel:2013rza,
    author = {Typel, S. and Oertel, M. and Kl{\"a}hn, T.},
    title = "{CompOSE CompStar online supernova equations of state harmonising the concert of nuclear physics and astrophysics compose.obspm.fr}",
    eprint = "1307.5715",
    archivePrefix = "arXiv",
    primaryClass = "astro-ph.SR",
    doi = "10.1134/S1063779615040061",
    journal = "Phys. Part. Nucl.",
    volume = "46",
    number = "4",
    pages = "633--664",
    year = "2015"
}

@article{CompOSECoreTeam:2022ddl,
    author = "Typel, S. and others",
    collaboration = "CompOSE Core Team",
    title = "{CompOSE Reference Manual}",
    eprint = "2203.03209",
    archivePrefix = "arXiv",
    primaryClass = "astro-ph.HE",
    doi = "10.1140/epja/s10050-022-00847-y",
    journal = "Eur. Phys. J. A",
    volume = "58",
    number = "11",
    pages = "221",
    year = "2022"
}

@article{Mroczek:2023zxo,
    author = "Mroczek, Debora and Miller, M. Coleman and Noronha-Hostler, Jacquelyn and Yunes, Nicolas",
    title = "{Nontrivial features in the speed of sound inside neutron stars}",
    eprint = "2309.02345",
    archivePrefix = "arXiv",
    primaryClass = "astro-ph.HE",
    doi = "10.1103/PhysRevD.110.123009",
    journal = "Phys. Rev. D",
    volume = "110",
    number = "12",
    pages = "123009",
    year = "2024"
}

@article{Loffler:2011ay,
    author = "Loffler, Frank and others",
    title = "{The Einstein Toolkit: A Community Computational Infrastructure for Relativistic Astrophysics}",
    eprint = "1111.3344",
    archivePrefix = "arXiv",
    primaryClass = "gr-qc",
    doi = "10.1088/0264-9381/29/11/115001",
    journal = "Class. Quant. Grav.",
    volume = "29",
    pages = "115001",
    year = "2012"
}

@article{Radice:2013xpa,
    author = "Radice, David and Rezzolla, Luciano and Galeazzi, Filippo",
    title = "{High-Order Fully General-Relativistic Hydrodynamics: new Approaches and Tests}",
    eprint = "1312.5004",
    archivePrefix = "arXiv",
    primaryClass = "gr-qc",
    doi = "10.1088/0264-9381/31/7/075012",
    journal = "Class. Quant. Grav.",
    volume = "31",
    pages = "075012",
    year = "2014"
}

@book{Baumgarte:2010,
  author        = "Baumgarte, Thomas W. and Shapiro, Stuart L.",
  title         = "{Numerical Relativity: Solving Einstein's Equations on the Computer}",
  publisher     = "{Cambridge University Press}",
  address       = "{Cambridge, UK}",
  year          = "2010",
  doi           = "10.1017/CBO9781139193344",
  isbn          = "978-0-521-51407-1"
}

@article{Dietrich:2018phi,
    author = {Dietrich, Tim and Radice, David and Bernuzzi, Sebastiano and Zappa, Francesco and Perego, Albino and Br{\"u}gmann, Bernd and Chaurasia, Swami Vivekanandji and Dudi, Reetika and Tichy, Wolfgang and Ujevic, Maximiliano},
    title = "{CoRe database of binary neutron star merger waveforms}",
    eprint = "1806.01625",
    archivePrefix = "arXiv",
    primaryClass = "gr-qc",
    doi = "10.1088/1361-6382/aaebc0",
    journal = "Class. Quant. Grav.",
    volume = "35",
    number = "24",
    pages = "24LT01",
    year = "2018"
}

@article{Fattoyev:2020cws,
    author = "Fattoyev, F. J. and Horowitz, C. J. and Piekarewicz, J. and Reed, Brendan",
    title = "{GW190814: Impact of a 2.6 solar mass neutron star on the nucleonic equations of state}",
    eprint = "2007.03799",
    archivePrefix = "arXiv",
    primaryClass = "nucl-th",
    doi = "10.1103/PhysRevC.102.065805",
    journal = "Phys. Rev. C",
    volume = "102",
    number = "6",
    pages = "065805",
    year = "2020"
}

@article{Banik:2014qja,
    author = "Banik, Sarmistha and Hempel, Matthias and Bandyopadhyay, Debades",
    title = "{New Hyperon Equations of State for Supernovae and Neutron Stars in Density-dependent Hadron Field Theory}",
    eprint = "1404.6173",
    archivePrefix = "arXiv",
    primaryClass = "astro-ph.HE",
    doi = "10.1088/0067-0049/214/2/22",
    journal = "Astrophys. J. Suppl.",
    volume = "214",
    number = "2",
    pages = "22",
    year = "2014"
}

@article{Steiner:2012rk,
    author = "Steiner, Andrew W. and Hempel, Matthias and Fischer, Tobias",
    title = "{Core-collapse supernova equations of state based on neutron star observations}",
    eprint = "1207.2184",
    archivePrefix = "arXiv",
    primaryClass = "astro-ph.SR",
    reportNumber = "INT-PUB-12-033",
    doi = "10.1088/0004-637X/774/1/17",
    journal = "Astrophys. J.",
    volume = "774",
    pages = "17",
    year = "2013"
}

@article{Lattimer:1991nc,
    author = "Lattimer, James M. and Swesty, F. Douglas",
    title = "{A Generalized equation of state for hot, dense matter}",
    doi = "10.1016/0375-9474(91)90452-C",
    journal = "Nucl. Phys. A",
    volume = "535",
    pages = "331--376",
    year = "1991"
}

@article{Bernuzzi:2014kca,
    author = "Bernuzzi, Sebastiano and Nagar, Alessandro and Balmelli, Simone and Dietrich, Tim and Ujevic, Maximiliano",
    title = "{Quasiuniversal properties of neutron star mergers}",
    eprint = "1402.6244",
    archivePrefix = "arXiv",
    primaryClass = "gr-qc",
    doi = "10.1103/PhysRevLett.112.201101",
    journal = "Phys. Rev. Lett.",
    volume = "112",
    pages = "201101",
    year = "2014"
}

@article{Bauswein:2010dn,
    author = "Bauswein, A. and Janka, H. -Th. and Oechslin, R.",
    title = "{Testing Approximations of Thermal Effects in Neutron Star Merger Simulations}",
    eprint = "1006.3315",
    archivePrefix = "arXiv",
    primaryClass = "astro-ph.SR",
    doi = "10.1103/PhysRevD.82.084043",
    journal = "Phys. Rev. D",
    volume = "82",
    pages = "084043",
    year = "2010"
}

@article{Fraga:2013qra,
    author = "Fraga, Eduardo S. and Kurkela, Aleksi and Vuorinen, Aleksi",
    title = "{Interacting quark matter equation of state for compact stars}",
    eprint = "1311.5154",
    archivePrefix = "arXiv",
    primaryClass = "nucl-th",
    reportNumber = "CERN-PH-TH-2013-269, HIP-2013-27-TH",
    doi = "10.1088/2041-8205/781/2/L25",
    journal = "Astrophys. J. Lett.",
    volume = "781",
    number = "2",
    pages = "L25",
    year = "2014"
}

@article{Gonzalez:2022mgo,
    author = "Gonzalez, Alejandra and others",
    title = "{Second release of the CoRe database of binary neutron star merger waveforms}",
    eprint = "2210.16366",
    archivePrefix = "arXiv",
    primaryClass = "gr-qc",
    doi = "10.1088/1361-6382/acc231",
    journal = "Class. Quant. Grav.",
    volume = "40",
    number = "8",
    pages = "085011",
    year = "2023"
}

@article{Weyhausen:2011cg,
    author = "Weyhausen, Andreas and Bernuzzi, Sebastiano and Hilditch, David",
    title = "{Constraint damping for the Z4c formulation of general relativity}",
    eprint = "1107.5539",
    archivePrefix = "arXiv",
    primaryClass = "gr-qc",
    doi = "10.1103/PhysRevD.85.024038",
    journal = "Phys. Rev. D",
    volume = "85",
    pages = "024038",
    year = "2012"
}

@article{Hinderer:2009ca,
    author = "Hinderer, Tanja and Lackey, Benjamin D. and Lang, Ryan N. and Read, Jocelyn S.",
    title = "{Tidal deformability of neutron stars with realistic equations of state and their gravitational wave signatures in binary inspiral}",
    eprint = "0911.3535",
    archivePrefix = "arXiv",
    primaryClass = "astro-ph.HE",
    doi = "10.1103/PhysRevD.81.123016",
    journal = "Phys. Rev. D",
    volume = "81",
    pages = "123016",
    year = "2010"
}

@article{Rosswog:2024vfe,
    author = "Rosswog, Stephan and Sarin, Nikhil and Nakhar, Ehud and Diener, Peter",
    title = "{Fast dynamic ejecta in neutron star mergers}",
    eprint = "2411.18813",
    archivePrefix = "arXiv",
    primaryClass = "astro-ph.HE",
    doi = "10.1093/mnras/staf324",
    journal = "Mon. Not. Roy. Astron. Soc.",
    volume = "538",
    number = "2",
    pages = "907--924",
    year = "2025"
}

@article{Cai:2025nxn,
    author = "Cai, Bao-Jun and Li, Bao-An",
    title = "{Novel scalings of neutron star properties from analyzing dimensionless Tolman{\textendash}Oppenheimer{\textendash}Volkoff equations}",
    eprint = "2501.18676",
    archivePrefix = "arXiv",
    primaryClass = "astro-ph.HE",
    doi = "10.1140/epja/s10050-025-01507-7",
    journal = "Eur. Phys. J. A",
    volume = "61",
    number = "3",
    pages = "55",
    year = "2025"
}

@Misc{EinsteinToolkit:2025_05,
  requested-for ={EinsteinToolkit},
  author       = {Maxwell Rizzo and Roland Haas and Steven R. Brandt and Zachariah Etienne and Deborah Ferguson and Lucas Timotheo Sanches and Bing-Jyun Tsao and Leonardo Werneck and David Boyer and Gabriele Bozzola and Cheng-Hsin Cheng and Samuel Cupp and Peter Diener and Terrence Pierre Jacques and Liwei Ji and Hayley Macpherson and Ivan Markin and Erik Schnetter and Wolfgang Tichy and Samuel Tootle and Yumeng Xu and Miguel Zilhão and Yosef Zlochower and Miguel Alcubierre and Daniela Alic and Gabrielle Allen and Marcus Ansorg and Federico G. Lopez Armengol and Maria Babiuc-Hamilton and Luca Baiotti and Werner Benger and Eloisa Bentivegna and Sebastiano Bernuzzi and Krishiv Bhatia and Tanja Bode and Brockton Brendal and Bernd Bruegmann and Manuela Campanelli and Michail Chabanov and Federico Cipolletta and Giovanni Corvino and Roberto De Pietri and Alexandru Dima and Harry Dimmelmeier and Jake Doherty and Rion Dooley and Nils Dorband and Matthew Elley and Yaakoub El Khamra and Lorenzo Ennoggi and Joshua Faber and Giuseppe Ficarra and Toni Font and Joachim Frieben and Bruno Giacomazzo and Tom Goodale and Carsten Gundlach and Ian Hawke and Scott Hawley and Ian Hinder and E. A. Huerta and Sascha Husa and Taishi Ikeda and Sai Iyer and Daniel Johnson and Abhishek V. Joshi and Jay Kalinani and Anuj Kankani and Wolfgang Kastaun and Thorsten Kellermann and Andrew Knapp and Michael Koppitz and Pablo Laguna and Gerd Lanferman and Paul Lasky and Frank Löffler and Joan Masso and Lars Menger and Andre Merzky and Jonah Maxwell Miller and Mark Miller and Philipp Moesta and Pedro Montero and Bruno Mundim and Patrick Nelson and Andrea Nerozzi and Scott C. Noble and Christian Ott and Ludwig Jens Papenfort and Ravi Paruchuri and Michal Pirog and Denis Pollney and Daniel Price and David Radice and Thomas Radke and Christian Reisswig and Luciano Rezzolla and Chloe B. Richards and David Rideout and Matei Ripeanu and Lorenzo Sala and Jascha A Schewtschenko and Bernard Schutz and Ed Seidel and Eric Seidel and John Shalf and Swapnil Shankar and Ken Sible and Ulrich Sperhake and Nikolaos Stergioulas and Wai-Mo Suen and Bela Szilagyi and Ryoji Takahashi and Michael Thomas and Jonathan Thornburg and Chi Tian and Malcolm Tobias and Aaryn Tonita and Paul Walker and Mew-Bing Wan and Barry Wardell and Helvi Witek and Burkhard Zink},
  title        = {The {E}instein {T}oolkit},
  month        = may,
  year         = 2025,
  publisher    = {Zenodo},
  version      = {The "Martin D. Kruskal" release, ET\_2025\_05},
  doi          = {10.5281/zenodo.15520463},
  url          = {https://doi.org/10.5281/zenodo.15520463},
}

@misc{arXiv:2606.21402,
      author       = "Shi Jian-Hao and Cai, Bao-Jun and Ma, Yu-Gang",
      title        = "A New Scaling of Neutron Star Tidal Deformability for Directly Probing the Core Equation of State",
      year         = "2026",
      eprint       = "2606.21402",
      archivePrefix= "arXiv",
      primaryClass = "astro-PH.HE"
}

@article{DelPozzo:2013ala,
    author = "Del Pozzo, Walter and Li, Tjonnie G. F. and Agathos, Michalis and Van Den Broeck, Chris and Vitale, Salvatore",
    title = "{Demonstrating the feasibility of probing the neutron star equation of state with second-generation gravitational wave detectors}",
    eprint = "1307.8338",
    archivePrefix = "arXiv",
    primaryClass = "gr-qc",
    doi = "10.1103/PhysRevLett.111.071101",
    journal = "Phys. Rev. Lett.",
    volume = "111",
    number = "7",
    pages = "071101",
    year = "2013"
}

@article{Chatziioannou:2020pqz,
    author = "Chatziioannou, Katerina",
    title = "{Neutron star tidal deformability and equation of state constraints}",
    eprint = "2006.03168",
    archivePrefix = "arXiv",
    primaryClass = "gr-qc",
    doi = "10.1007/s10714-020-02754-3",
    journal = "Gen. Rel. Grav.",
    volume = "52",
    number = "11",
    pages = "109",
    year = "2020"
}

@article{Lattimer:2012nd,
  author = "Lattimer, James M.",
  title = "{The nuclear equation of state and neutron star masses}",
  journal = "Ann. Rev. Nucl. Part. Sci.",
  volume = "62",
  pages = "485-515",
  year = "2012"
}

@article{Oertel:2016bki,
  author = "Oertel, Micaela and Hempel, Matthias and Klaehn, Thomas and Typel, Stefan",
  title = "{Equations of state for supernovae and compact stars}",
  journal = "Rev. Mod. Phys.",
  volume = "89",
  pages = "015007",
  year = "2017"
}

@article{Hulse:1974eb,
  author = "Hulse, R. A. and Taylor, J. H.",
  title = "{Discovery of a pulsar in a binary system}",
  journal = "Astrophys. J. Lett.",
  volume = "195",
  pages = "L51-L53",
  year = "1975"
}

@article{Taylor:1982zz,
  author = "Taylor, J. H. and Weisberg, J. M.",
  title = "{A new test of general relativity: Gravitational radiation and the binary pulsar PSR 1913+16}",
  journal = "Astrophys. J.",
  volume = "253",
  pages = "908-920",
  year = "1982"
}

@article{Abbott:2017vwq,
  author = "{LIGO Scientific Collaboration} and {Virgo Collaboration}",
  title = "{GW170817: Observation of gravitational waves from a binary neutron star inspiral}",
  journal = "Phys. Rev. Lett.",
  volume = "119",
  pages = "161101",
  year = "2017"
}

@article{Metzger:2016pju,
  author = "Metzger, Brian D.",
  title = "{Kilonovae}",
  journal = "Living Rev. Rel.",
  volume = "23",
  year = "2017"
}

@article{Kasen:2017sxr,
  author = "Kasen, Daniel and Metzger, Brian and Barnes, Jennifer and Quataert, Eliot and Ramirez-Ruiz, Enrico",
  title = "{Origin of the heavy elements in binary neutron-star mergers from a gravitational-wave event}",
  journal = "Nature",
  volume = "551",
  pages = "80-84",
  year = "2017"
}

@article{Flanagan:2007ix,
  author = "Flanagan, Eanna E. and Hinderer, Tanja",
  title = "{Constraining neutron star tidal Love numbers with gravitational wave detectors}",
  journal = "Phys. Rev. D",
  volume = "77",
  pages = "021502",
  year = "2008"
}

@article{Hinderer:2007mb,
  author = "Hinderer, Tanja",
  title = "{Tidal Love numbers of neutron stars}",
  journal = "Astrophys. J.",
  volume = "677",
  pages = "1216-1220",
  year = "2008"
}

@article{Bauswein:2011tp,
  author = "Bauswein, Andreas and Janka, Hans-Thomas",
  title = "{Measuring neutron-star properties via gravitational waves from binary mergers}",
  journal = "Phys. Rev. Lett.",
  volume = "108",
  pages = "011101",
  year = "2012"
}

@article{Hotokezaka:2013iia,
  author = "Hotokezaka, Kenta and others",
  title = "{Remnant massive neutron stars of binary neutron star mergers: Evolution process and gravitational waveform}",
  journal = "Phys. Rev. D",
  volume = "88",
  pages = "044026",
  year = "2013"
}

@article{Bauswein:2013jpa,
  author = "Bauswein, Andreas and Baumgarte, Thomas W. and Janka, Hans-Thomas",
  title = "{Prompt merger collapse and the maximum mass of neutron stars}",
  journal = "Phys. Rev. Lett.",
  volume = "111",
  pages = "131101",
  year = "2013"
}

@article{Koeppel:2019pys,
  author = "K{\"o}ppel, S. and Bovard, L. and Rezzolla, L.",
  title = "{A General-Relativistic Determination of the Threshold Mass to Prompt Collapse in Binary Neutron Sar Mergers}",
  journal = "Astrophys. J. Lett.",
  volume = "872",
  pages = "L16",
  year = "2019"
}

@article{Demorest:2010bx,
  author = "Demorest, P. B. and Pennucci, T. and Ransom, S. M. and Roberts, M. S. E. and Hessels, J. W. T.",
  title = "{Shapiro Delay Measurement of A Two Solar Mass Neutron Star}",
  eprint = "1010.5788",
  archivePrefix = "arXiv",
  primaryClass = "astro-ph.HE",
  doi = "10.1038/nature09466",
  journal = "Nature",
  volume = "467",
  pages = "1081--1083",
  year = "2010"
}

@article{Antoniadis:2013pzd,
  author = "Antoniadis, J. and Freire, P. C. C. and Wex, N. and Tauris, T. M. and Lynch, R. S. and others",
  title = "{A Massive Pulsar in a Compact Relativistic Binary}",
  eprint = "1304.6875",
  archivePrefix = "arXiv",
  primaryClass = "astro-ph.HE",
  doi = "10.1126/science.1233232",
  journal = "Science",
  volume = "340",
  pages = "6131",
  year = "2013"
}

@article{Fonseca:2021wxt,
  author = "Fonseca, E. and Cromartie, H. T. and Pennucci, T. T. and Ray, P. S. and Kirichenko, A. Y. and others",
  title = "{Refined Mass and Geometric Measurements of the High-Mass PSR J0740+6620}",
  eprint = "2104.00880",
  archivePrefix = "arXiv",
  primaryClass = "astro-ph.HE",
  doi = "10.3847/1538-4357/ac03b8",
  journal = "Astrophys. J. Lett.",
  volume = "915",
  number = "1",
  pages = "L12",
  year = "2021"
}

@article{Riley:2019yda,
  author = "Riley, T. E. and Watts, A. L. and Bogdanov, S. and Ray, P. S. and Ludlam, R. M. and others",
  title = "{A NICER View of PSR J0030+0451: Millisecond Pulsar Parameter Estimation}",
  eprint = "1912.05702",
  archivePrefix = "arXiv",
  primaryClass = "astro-ph.HE",
  doi = "10.3847/2041-8213/ab481c",
  journal = "Astrophys. J. Lett.",
  volume = "887",
  number = "1",
  pages = "L21",
  year = "2019"
}

@article{Miller:2019cac,
  author = "Miller, M. C. and Lamb, F. K. and Dittmann, A. J. and Bogdanov, S. and Arzoumanian, Z. and others",
  title = "{PSR J0030+0451 Mass and Radius from NICER Data and Implications for the Properties of Neutron Star Matter}",
  eprint = "1912.05705",
  archivePrefix = "arXiv",
  primaryClass = "astro-ph.HE",
  doi = "10.3847/2041-8213/ab50c5",
  journal = "Astrophys. J. Lett.",
  volume = "887",
  number = "1",
  pages = "L24",
  year = "2019"
}

@article{Riley:2021pdl,
  author = "Riley, T. E. and Watts, A. L. and Ray, P. S. and Morsink, S. M. and Bilous, A. V. and others",
  title = "{A NICER View of PSR J0740+6620 Informing the Neutron Star Equation of State}",
  eprint = "2105.06980",
  archivePrefix = "arXiv",
  primaryClass = "astro-ph.HE",
  doi = "10.3847/2041-8213/ac0a81",
  journal = "Astrophys. J. Lett.",
  volume = "918",
  number = "2",
  pages = "L27",
  year = "2021"
}

@article{Miller:2021qha,
  author = "Miller, M. C. and Lamb, F. K. and Dittmann, A. J. and Bogdanov, S. and Arzoumanian, Z. and others",
  title = "{The Radius of PSR J0740+6620 from NICER and XMM-Newton Data}",
  eprint = "2105.06979",
  archivePrefix = "arXiv",
  primaryClass = "astro-ph.HE",
  doi = "10.3847/2041-8213/ac089b",
  journal = "Astrophys. J. Lett.",
  volume = "918",
  number = "2",
  pages = "L28",
  year = "2021"
}

@article{Abbott:2018exr,
  author = "Abbott, B. P. and others",
  collaboration = "LIGO Scientific, Virgo",
  title = "{Properties of the Binary Neutron Star Merger GW170817}",
  eprint = "1805.11579",
  archivePrefix = "arXiv",
  primaryClass = "gr-qc",
  doi = "10.1103/PhysRevX.9.011001",
  journal = "Phys. Rev. X",
  volume = "9",
  number = "1",
  pages = "011001",
  year = "2019"
}

@article{Bedaque:2014sqa,
  author = "Bedaque, Paulo and Steiner, Andrew W.",
  title = "{Sound Velocity Bound and Neutron Stars}",
  eprint = "1408.5116",
  archivePrefix = "arXiv",
  primaryClass = "nucl-th",
  doi = "10.1103/PhysRevLett.114.031103",
  journal = "Phys. Rev. Lett.",
  volume = "114",
  number = "3",
  pages = "031103",
  year = "2015"
}

@article{Tews:2018iwm,
  author = "Tews, Ingo and Margueron, Jerome and Reddy, Sanjay",
  title = "{A critical examination of constraints on the equation of state of dense matter obtained from GW170817}",
  eprint = "1804.02783",
  archivePrefix = "arXiv",
  primaryClass = "nucl-th",
  doi = "10.1103/PhysRevC.98.045804",
  journal = "Phys. Rev. C",
  volume = "98",
  number = "4",
  pages = "045804",
  year = "2018"
}

@article{Annala:2019puf,
  author = "Annala, Eemeli and Gorda, Tyler and Kurkela, Aleksi and Nattila, Joonas and Vuorinen, Aleksi",
  title = "{Evidence for Quark-Matter Cores in Massive Neutron Stars}",
  eprint = "1903.09121",
  archivePrefix = "arXiv",
  primaryClass = "astro-ph.HE",
  doi = "10.1038/s41586-020-2303-2",
  journal = "Nature Phys.",
  volume = "16",
  number = "9",
  pages = "907--910",
  year = "2020"
}

@article{Ecker:2022xxx,
  author = "Ecker, Christian and Rezzolla, Luciano",
  title = "{A General, Scale-Independent Description of the Sound Speed in Neutron Stars}",
  eprint = "2207.04417",
  archivePrefix = "arXiv",
  primaryClass = "astro-ph.HE",
  doi = "10.3847/2041-8213/ac8674",
  journal = "Astrophys. J. Lett.",
  volume = "939",
  number = "2",
  pages = "L35",
  year = "2022"
}

@article{Hebeler:2010xb,
  author = {Hebeler, K. and Schwenk, A.},
  title = {Chiral three-nucleon forces and neutron matter},
  journal = {Phys. Rev. C},
  volume = {82},
  pages = {014314},
  year = {2010}
}

@article{Drischler:2016djf,
  author = {Drischler, Christian and Carbone, Arianna and Soma, Takashi and Schwenk, Achim},
  title = {Neutron Matter from chiral two- and three-nucleon calculations up to $N^3$L0},
  journal = {Phys. Rev. C},
  volume = {94},
  pages = {054307},
  year = {2016}
}

@article{Komoltsev:2022tsh,
  author = {Komoltsev, Oleg and Kurkela, Aleksi},
  title = {How perturbative QCD constrains the Equation of State at neutron-star densities},
  journal = {Phys. Rev. Lett.},
  volume = {128},
  pages = {202701},
  year = {2022}
}

@article{Annala:2022sdu,
  author = {Annala, Eemeli and Gorda, Tyler and Hirvonen, Joonas and Komoltsev, Oleg and Kurkela, Aleksi and Naettilae, Joonas and Vuorinen, Aleksi},
  title = {Strongly interacting matter exhibits deconfined behavior in massive neutron stars},
  journal = {Nat. Commun.},
  volume = {14},
  pages = {8451},
  year = {2023}
}

@article{Hebeler:2009iv,
    author = "Hebeler, K. and Schwenk, A.",
    title = "{Chiral three-nucleon forces and neutron matter}",
    eprint = "0911.0483",
    archivePrefix = "arXiv",
    primaryClass = "nucl-th",
    doi = "10.1103/PhysRevC.82.014314",
    journal = "Phys. Rev. C",
    volume = "82",
    pages = "014314",
    year = "2010"
}

@article{Tews:2012fj,
    author = "Tews, Ingo and Kruger, Thomas and Hebeler, Kai and Schwenk, Achim",
    title = "{Neutron matter at next-to-next-to-next-to-leading order in chiral effective field theory}",
    eprint = "1206.0025",
    archivePrefix = "arXiv",
    primaryClass = "nucl-th",
    doi = "10.1103/PhysRevLett.110.032504",
    journal = "Phys. Rev. Lett.",
    volume = "110",
    pages = "032504",
    year = "2013"
}

@article{Drischler:2019wtt,
    author = "Drischler, Christian and Melendez, J. A. and Furnstahl, R. J. and Phillips, D. R.",
    title = "{Quantifying uncertainties and correlations in the nuclear-matter equation of state}",
    eprint = "2004.07805",
    archivePrefix = "arXiv",
    primaryClass = "nucl-th",
    doi = "10.1103/PhysRevLett.122.042501",
    journal = "Phys. Rev. C.",
    volume = "102",
    pages = "042501",
    year = "2020"
}

@article{Drischler:2020vpf,
    author = "Drischler, Christian and Som\`a, Vittorio and Schwenk, Achim",
    title = "{Microscopic calculations and energy expansions for neutron-rich matter}",
    eprint = "1310.5627",
    archivePrefix = "arXiv",
    primaryClass = "nucl-th",
    doi = "10.1103/PhysRevC.102.054315",
    journal = "Phys. Rev. C",
    volume = "89",
    pages = "025806",
    year = "2014"
}

@article{Huth:2021xxh,
    author = "Huth, Sabrina and Pang, Peter and Tews, Ingo and Dietrich, Tim and Le Fevre, Arnaud and Schwenk, Achim and Trautmann, Wolfgang and Agarwal, Kshitij and Bulla, Mattia and Coughlin, Micheal and Van Den Broeck, Chris",
    title = "{Constraints on neutron-star matter from microscopic and macroscopic collisions}",
    eprint = "2107.06229",
    archivePrefix = "arXiv",
    primaryClass = "nucl-th",
    doi = "10.1038/s41586-022-04750-w",
    journal = "Nature Phys.",
    volume = "606",
    pages = "276--280",
    year = "2022"
}

@article{Gandolfi:2011xu,
    author = "Gandolfi, Stefano and Carlson, Joseph and Reddy, Sanjay",
    title = "{The maximum mass and radius of neutron stars and the nuclear symmetry energy}",
    eprint = "1101.1921",
    archivePrefix = "arXiv",
    primaryClass = "nucl-th",
    doi = "10.1103/PhysRevC.85.032801",
    journal = "Phys. Rev. C",
    volume = "85",
    number = "3",
    pages = "032801",
    year = "2012"
}

@article{Gezerlis:2013ipa,
    author = "Gezerlis, Alexandros and Tews, Ingo and Epelbaum, Evgeny and Gandolfi, Stefano and Hebeler, Kai and Nogga, Andreas and Schwenk, Achim",
    title = "{Quantum Monte Carlo Calculations with Chiral Effective Field Theory Interactions}",
    eprint = "1303.6243",
    archivePrefix = "arXiv",
    primaryClass = "nucl-th",
    doi = "10.1103/PhysRevLett.111.032501",
    journal = "Phys. Rev. Lett.",
    volume = "111",
    number = "3",
    pages = "032501",
    year = "2013"
}

@article{Lynn:2015jua,
    author = "Lynn, J. E. and Tews, I. and Carlson, J. and Gandolfi, S. and Gezerlis, A. and Schmidt, K. E. and Schwenk, A.",
    title = "{Chiral Three-Nucleon Interactions in Light Nuclei, Neutron-$\alpha$ Scattering, and Neutron Matter}",
    eprint = "1509.03470",
    archivePrefix = "arXiv",
    primaryClass = "nucl-th",
    doi = "10.1103/PhysRevLett.116.062501",
    journal = "Phys. Rev. Lett.",
    volume = "116",
    number = "6",
    pages = "062501",
    year = "2016"
}

@article{Lonardoni:2019ypa,
    author = "Lonardoni, D. and Gandolfi, S. and Lynn, J.E. and Petrie, C. and Carlson, J. and Schmidt, K.E. and Schwenk, A.",
    title = "{Auxiliary field Diffusion monte carlo calculations of light and medium mass nuclei with local chiral interactions}",
    eprint = "1904.08050",
    archivePrefix = "arXiv",
    primaryClass = "nucl-th",
    doi = "10.1103/PhysRevC.97.044318",
    journal = "Phys. Rev. C",
    volume = "97",
    number = "4",
    pages = "044318",
    year = "2018"
}

@article{Kruger:2013kua,
    author = {Kr{\"u}ger, T. and Tews, I. and Hebeler, K. and Schwenk, A.},
    title = "{Neutron matter from chiral effective field theory interactions}",
    eprint = "1304.2212",
    archivePrefix = "arXiv",
    primaryClass = "nucl-th",
    doi = "10.1103/PhysRevC.88.025802",
    journal = "Phys. Rev. C",
    volume = "88",
    pages = "025802",
    year = "2013"
}
\end{document}